\documentclass[a4paper,12pt]{article}
\pdfoutput=1 % if your are submitting a pdflatex (i.e. if you have
\usepackage{jheppub} % for details on the use of the package, please
\usepackage[T1]{fontenc} % if needed

\title{\boldmath Factorization of the 3d superconformal index with an adjoint matter}

\author[a]{Chiung Hwang,}
\author[a,b]{Jaemo Park,}

\affiliation[a]{Department of Physics, POSTECH,\\Pohang 790-784, Korea}
\affiliation[b]{Postech Center for Theoretical Physics (PCTP), POSTECH,\\Pohang 790-784, Korea}

\emailAdd{c\_hwang@postech.ac.kr}
\emailAdd{jaemo@postech.ac.kr}

\abstract{ We work out the factorization of the 3d superconformal index for $\mathcal N = 2$ $U(N_c)$ gauge theory with
 one adjoint chiral multiplet as well as $N_f$ fundamental, $N_a$ anti-fundamental chiral multiplets. Using the factorization,
 one can prove the Seiberg-like duality for $\mathcal N = 4$ $U(N_c)$ theory with $N_f$ hypermultiplets at the index level. We explicitly
 show that monopole operators violating unitarity bound in a bad theory are mapped to free hypermultiplets in the dual side.
 For $\mathcal N = 2$ $U(N_c)$ theory with one adjoint matter $X$,  $N_f$ fundamental, $N_a$ anti-fundamental chiral multiplets with superpotential $W = \mathrm{tr} X^{n+1}$,
 we work out Seiberg-like duality for this theory. The index computation provides combinatorial identities for a dual pair, which we carry out
 intensive numerical checks. }

\begin{document}
\maketitle
\flushbottom

\section{Introduction}
The 3d superconformal index \cite{Bhattacharya:2008zy,Bhattacharya:2008bja,
Kim:2009wb, Imamura:2011su} has been used to probe the various dualities for $\mathcal N = 2$ supersymmetric gauge theories \cite{Krattenthaler:2011da,Kapustin:2011jm,Bashkirov:2011vy,Hwang:2011qt,Hwang:2011ht,Kapustin:2011vz,Dimofte:2011py,Kim:2013cma,Park:2013wta,Aharony:2013dha}.
One of the interesting feature of the 3d index is its factorization into vortex and antivoretx partition function \cite{Beem:2012mb,Hwang:2012jh}.
Schematically
\begin{equation}
I(z,\bar{z})= Z_\text{vortex}(z) Z_\text{antivortex} (\bar{z})= |Z_\text{vortex}(z)|^2
\end{equation}
where $z$ traces the vortex number and $ \bar{z}$ traces antivortex number.
This is the 3d analogue of the 2d conformal blocks.
Previously it was shown that such factorization holds for $\mathcal N = 2$ $U(N_c)_\kappa$ gauge theory with $N_f$ fundamental chiral multiplets and
$N_a$ anti-fundamental chiral multiplets with  Chern-Simons (CS) levels satisfying $|\kappa| \leq \frac{|N_f-N_a|}{2}$ \cite{Benini:2013yva}.
Once we have the factorized form of the 3d index, the proof of the duality is reduced to the proof of some combinatorial
identities at the index level \cite{Hwang:2012jh}. Hence the structure of the duality is much more transparent once such factorized index is available.

We extend the previous proof of the factorization into $\mathcal N = 2$ $U(N_c)$ theory with one  adjoint chiral multiplets as well as fundamental and anti-fundamental
matters. Once we obtain the factorization, we apply the result to two cases.
As the 1st application, we consider the Seiberg-like duality for $\mathcal N = 4$ $U(N_c)$ gauge theory with $N_f$ hypermultiplets \cite{Kim:2012uz,Yaakov:2013fza,Gaiotto:2013bwa}.
For $N_c \leq N_f < 2N_c-1$ the theory is called ``bad'' since it contains the monopole operators whose conformal dimension violates the unitarity bound.
The fate of such monopole operators in IR is interesting. For the  case in hand, such monopole operators are flowing to free hypermultiplets in IR
and there are some evidences put forward at \cite{Kim:2012uz,Yaakov:2013fza,Gaiotto:2013bwa,Razamat:2014pta}.
Here using the factorization of the index, we explicitly prove the conjectured dualities at the index level.
Along with that, we can show how pathological $R$-charges in UV can mix with accidental global symmetries in IR to give the correct $R$-charges in IR.
As the second application, we consider the dualities of $\mathcal N = 2$ $U(N_c)$ gauge theories with one adjoint $X$, fundamental and anti-fundamental matters \cite{Kim:2013cma}.
In this case we need to introduce the superpotential $W=\mathrm{tr} X^n$ to have the dual pair. We generalize the previous work \cite{Kim:2013cma}
 to the cases where one can have different number of chiral multiplets and anti-chiral multiplets, which we call chiral-like theories, with Chern-Simons terms. For the above theory and the conjectured dual pair, we
work out the factorized index and the equality of the proof is reduced to the proof of combinatorial identities, which we carry out the intensive
numerical checks. We also consider the cases where some monopole operators violate the unitarity bounds. For some cases we understand the fate of
such operators in IR. However the understanding is rather limited compared with $\mathcal N = 4$ cases.

The content of the paper is as follows. In section \ref{sec:factorization} we work out the factorization of $U(N_c)$ theory with one adjoint, fundamental and antifundamental
matters. In section \ref{sec:N=4}, we consider $\mathcal N = 4$ Seiberg-like dualities for $\mathcal N = 4$ $U(N_c)$ theory with fundamental hypermultiplets. The equality of the index for such
dual pairs are explicitly proved. Meanwhile we show that the monopole operators which violate the unitarity bound are turned into free hypermultiplets and
are decoupled. In section \ref{sec:N=2}, we consider the $\mathcal N = 2$  duality for $U(N_c)$ theory with one adjoint, fundamental and anti-fundamental matters using the factorized index. Technical details are relegated to appendices.

\section{Factorization with an adjoint matter} \label{sec:factorization}
In this section we examine the factorization of the superconformal index of the 3d $\mathcal N = 2$ $U(N_c)_\kappa$ gauge theory with $N_f$ fundamental, $N_a$ antifundamental, one adjoint matters and without superpotential, where $\kappa$ is the CS level. We focus on the values of $\kappa$ satisfying the condition $|\kappa| \leq \frac{|N_f-N_a|}{2}$ due to a technical issue we will explain. The matters are denoted by chiral multiplets $Q_a,\tilde Q^{\tilde b},X$, where $X$ denotes the adjoint chiral multiplet. The theory has  the global symmetry $U(1)_R \times SU(N_f) \times SU(N_a)  \times U(1)_A \times U(1)_X \times U(1)_T$, where $U(1)_R$ denotes the $R$ symmetry. The global charges of each chiral multiplet are summarized in table~\ref{tab:global charges}. The meaning of the global charges should be obvious from the table~\ref{tab:global charges}. In particular $U(1)_T$ is the topological
charge,  which monopole operators carry.
\begin{table}[tbp]
\centering
\begin{tabular}{|c|cccccc|}
\hline
 & $U(1)_R$ & $SU(N_f)$ & $SU(N_a)$ & $U(1)_A$ & $U(1)_X$ & $U(1)_T$ \\
\hline
$Q$ & $\Delta_Q$ & $\overline{\mathbf N_f}$ & $\mathbf 1$ & 1 & 0 & 0 \\
$\tilde Q$ & $\Delta_Q$ & $\mathbf 1$ & $\mathbf N_a$ & 1 & 0 & 0 \\
$X$ & $\Delta_X$ & $\mathbf 1$ & $\mathbf 1$ & 0 & 1 & 0 \\
\hline
$V_{i,\pm}$ & $\substack{\frac{1}{2} (1-\Delta_Q) (N_f+N_a)\\-\Delta_X (N_c-1-i)}$ & $\mathbf 1$ & $\mathbf 1$ & $-\frac{N_f+N_a}{2}$ & $-N_c+1+i$ & $\pm1$ \\
\hline
\end{tabular}
\caption{\label{tab:global charges} The global symmetry charges of chiral operators. The gauge invarint monopole operators $V_{i,\pm}$ appear when $\kappa\pm\frac{N_f-N_a}{2} = 0$.}
\end{table}
Compared to the theory without the adjoint matter, we have additional $U(1)_X$ symmetry, which rotates the phase of the adjoint chiral multiplet $X$. In addition, there are BPS bare monopole states labeled by the GNO \cite{Goddard:1976qe} charge lattice:
\begin{align*}
\left|m_1,\ldots,m_{N_c}\right>.
\end{align*}
If $\kappa\pm\frac{N_f-N_a}{2} = 0$, the bare monopole state $\left|\pm1,0,\ldots,0\right>$ is gauge invariant and corresponds to the gauge invariant chiral operator $V_{0,\pm}$. In addition, it can be dressed by the adjoint matter $X$ such that it corresponds to another chiral operator $V_{i,\pm} \sim \mathrm{tr} X^i \left|\pm1,0,\ldots,0\right>$ for $i = 0,\ldots,N_c-1$ \cite{Kim:2013cma,Cremonesi:2013lqa}. The global charges of those monopole operators appear also in table \ref{tab:global charges}. When one deals with the theory with the suprepotential, the only role of the superpotential in computing the superconformal index is just constraining the global charges consistent with the superpotential. Therefore, when we deal with a theory with a nonvanishing superpotential, we only need to impose proper global charges in the index computation.

Let us recall the definition of the superconformal index of the 3d $\mathcal N = 2$ theory. The bosonic subgroup of the 3d $\mathcal N = 2$ superconformal group is $SO(2,3) \times SO(2)$ whose three Cartan elements are denoted by $E,j$ and $R$. Then the superconformal index is defined by \cite{Bhattacharya:2008zy,Bhattacharya:2008bja}
\begin{align} \label{eq:index def}
I\left(x,t_i\right) = \mathrm{Tr} (-1)^F \exp(-\beta'\{Q,S\}) x^{E+j} \left(\prod_i t_i^{F_i}\right)
\end{align}
where $Q$ is a supercharge of quantum numbers $E = \frac{1}{2},j = -\frac{1}{2}$ and $R = 1$, while $S = Q^\dagger$. $x$ is the fugacity for $E+j$ and $t_i$'s are additional fugacities for global symmetries of the theory. The trace is taken over the Hilbert space of the SCFT on $\mathbb R \times S^2$, or equivalently over the space of local gauge invariant operators on $\mathbb R^3$. As usual, only the BPS states, which saturate the inequality
\begin{align}
\{Q,S\} = E-R-j \geq 0,
\end{align}
contribute to the index. Therefore, the index is independent of the parameter $\beta'$.

The matrix integral formula for the superconformal index of a $U(N_c)$ theory is given by \cite{Kim:2009wb,Imamura:2011su}
\begin{align}
\begin{aligned} \label{eq:matrix integral}
&\quad I(x,t,\tilde t,\tau,\upsilon,w) \\
&=\sum_{m\in\mathbb Z^{N_c}/S_{N_c}} \frac{1}{|\mathcal W_m|} \oint_{|z_j| = 1} \left(\prod_{j=1}^{N_c} \frac{dz_j}{2 \pi i z_j}\right) e^{-S_\text{CS}(z,m)} w^{\sum_j m_j} Z_\text{vector}(x,z,m) Z_\text{chiral}(x,t,\tilde t,\tau,\upsilon,z,m).
\end{aligned}
\end{align}
Here $\left|\mathcal W_m\right|$ is the Weyl group order of the residual gauge group left unbroken by the magnetic flux $m$. And  $z_j$'s are gauge holonomies along the time circle. $e^{-S_\text{CS}(z,m)}$ is the classical contribution of the CS term, which is written as
\begin{align}
e^{-S_\text{CS}(z,m)} = \prod_{j = 1}^{N_c} (-z_j)^{-\kappa m_j}.
\end{align}
$w$ is the fugacity for the $U(1)_T$ topological symmetry, whose conserved charge is given by $T = \sum_j m_j$. If we regard the symmetry as weakly gauged, $w$ corresponds to the background holonomy of the $U(1)_T$ symmetry. Therefore, the term $w^{\sum_j m_j}$ comes from  a mixed Chern-Simons term. $Z_\text{vector}$ and $Z_\text{chiral}$ are the 1-loop determinant contribution of the vector multiplet and the chiral multiplets respectively. The vector multiplet contribution is given by
\begin{align}
Z_\text{vector}(x,z,m) = \prod_{\substack{i,j = 1 \\ (i\neq j)}}^{N_c} x^{-|m_i-m_j|/2} \left(1-z_i z_j^{-1} x^{|m_i-m_j|}\right)
\end{align}
while the chiral multiplet contribution is given by
\begin{align}
Z_\text{chiral}(x,t,\tilde t,\tau,\upsilon,z,m) = Z_X(x,\upsilon,z,m) \left(\prod_{a = 1}^{N_f} Z_{Q_a}(x,t,\tau,z,m)\right) \left(\prod_{a = 1}^{N_a} Z_{\tilde Q^{\tilde b}}(x,\tilde t,\tau,z,m)\right)
\end{align}
where\footnote{$(a;q)_n$ is the $q$-Pochhammer symbol defined by
\begin{align}
(a;q)_n = \prod_{k = 0}^{n-1} \left(1-a q^k\right)
\end{align}}
\begin{align}
Z_{X}(x,\upsilon,z,m) &= \prod_{i,j = 1}^{N_c} \left(x^{\Delta_X-1} \upsilon\right)^{-|m_i-m_j|/2} \frac{\left(z_i^{-1} z_j \upsilon^{-1} x^{2-\Delta_X+|m_i-m_j|};x^2\right)_\infty}{\left(z_i z_j^{-1} \upsilon x^{\Delta_X+|m_i-m_j|};x^2\right)_\infty}, \\
Z_{Q_a}(x,t,\tau,z,m) &= \prod_{j = 1}^{N_c} \left(x^{\Delta_Q-1} (-z_j) t_a \tau\right)^{-|m_j|/2} \frac{\left(z_j^{-1} t_a^{-1} \tau^{-1} x^{2-\Delta_Q+|m_j|};x^2\right)_\infty}{\left(z_j t_a \tau x^{\Delta_Q+|m_j|};x^2\right)_\infty}, \\
Z_{\tilde Q^{\tilde b}}(x,\tilde t,\tau,z,m) &= \prod_{j = 1}^{N_c} \left(x^{\Delta_Q-1} (-z_j)^{-1} \tilde t_a \tau\right)^{-|m_j|/2} \frac{\left(z_j \tilde t_a^{-1} \tau^{-1} x^{2-\Delta_Q+|m_j|};x^2\right)_\infty}{\left(z_j^{-1} \tilde t_a \tau x^{\Delta_Q+|m_j|};x^2\right)_\infty}.
\end{align}
They are the contribution of the adjoint, fundamental and antifundamental chiral multiplets respectively. $t,\tilde t,\tau,\upsilon$ are the fugacities, or the background holonomies, for the global symmetry $SU(N_f) \times SU(N_a) \times U(1)_A \times U(1)_X$. We assume $|t| = |\tilde t| = 1$. In addition, $x$, the fugacity for $R+j$, is assumed to have the norm smaller than 1, $|x| < 1$. We will set $\Delta_Q = \Delta_X = 0$ for simplicity.\footnote{As defined at the table 1, $\Delta_Q ,\Delta_X $
are the $R$-charge of the fundamental and adjoint chiral multiplet respectively. Due to the BPS condition, this is equal to the conformal dimension of the multiplet.} They can have the different value $\Delta_Q, \Delta_X $ by shifting $\tau \rightarrow \tau x^{\Delta_Q}$ and $\upsilon \rightarrow \upsilon x^{\Delta_X}$. Because of the possibility of the shift, we will regard $|\tau|$ and $|\upsilon|$ as less than 1. When the theory has superpotential, we need to impose additional relations to the various fugacities, as we will see later. We also have included the phase factor $(-1)^{-\kappa \sum_j m_j-\frac{N_f-N_a}{2} \sum_j |m_j|}$, which is originated from the definition of the fermionic number operator $F$ in \eqref{eq:index def} \cite{Dimofte:2011py}. We use $F = 2 j+e \cdot m$ where $e$ and $m$ are electric charge and magnetic flux. One can introduce magnetic fluxes of background gauge fields coupled to the global symmetries \cite{Kapustin:2011jm}. However, we turn them off for simplicity.

The contour integral is iteratively evaluated for each $z_j$ along the unit circle, $|z_j| = 1$. Applying the residue theorem, we choose the poles inside the unit circle or alternatively choose the poles outside the unit circle with opposite sign because the  sum total of the residues is zero. For a technical reason, we consider $|\kappa| \leq \frac{|N_f-N_a|}{2}$ case, which is called ``maximally chiral'' in \cite{Benini:2011mf}.  If $N_f > N_a$, it is convenient to take the outside poles because there is no pole at infinity. If $N_f = N_a$, although there may exist poles both at the origin and at infinity, one can show that the residue at each pole vanishes \cite{Hwang:2012jh,Benini:2013yva}. On the other hand, for $|\kappa| > \frac{|N_f-N_a|}{2}$ case, which is called ``minimally chiral'' in \cite{Benini:2011mf}, both residues at the origin and at infinity do not vanish. In that case factorization of the superconformal index is not clear \cite{Benini:2013yva}. Thus, assuming $N_f \geq N_a$ we are going to take the poles outside the unit circle. The relevant poles are as follows:
\begin{align} \label{eq:poles}
z_j = \left\{\begin{array}{lc}
t_{a_j}^{-1} \tau^{-1} x^{-|m_j|-2 k_j}, & 1 \leq a_j \leq N_f, \quad k_j \geq 0 \\
z_i \upsilon^{-1} x^{-|m_j-m_i|-2 k_j}, & 1 \leq i (\neq j) \leq N_c, \quad k_j \geq 0
\end{array}\right.
\end{align}
where for each $j = 1,\ldots,N_c$, $z_j$ takes either the value in the first line with a choice of $a_j$,$k_j$ or the value in the second line with a choice of $i$ and $k_j$.
%We exclude the choices yielding a pole at the origin or infinity because their residues vanish. For example, a choice that every $z_j$ takes the second type of a pole is excluded.
Note that the first type of poles comes from the fundamental matter contribution while the second type of poles comes from the adjoint matter contribution. 
If we carry out the unit contour integration of $z_i$ by picking up outside poles, all poles of eq. (\ref{eq:poles}) are lying outside the unit circle 
irrespective of the value of $z_i$. One can have poles accidentally located outside  the unit circle depending on the value of  $z_i$.
For example the pole can have  the form $z_1=p, z_2=z_1 r^{-1} \cdots$ with $|p|>|r|>1$. This is not the type of eq. (\ref{eq:poles}). However if we sequentially
integrate over $z_1, z_2$, then integration over $z_2$ picks up the residue $z_2=p r^{-1}$, which is outside the unit circle of $z_2$ since $z_1$ takes the
specific value $p$. One can show that there is always  the cancelation of the residues for such accidentally picked-up poles. Thus we can consider the poles
specified at eq.  (\ref{eq:poles}).

\begin{figure}[tbp]
\centering
\includegraphics[width=.4\textwidth]{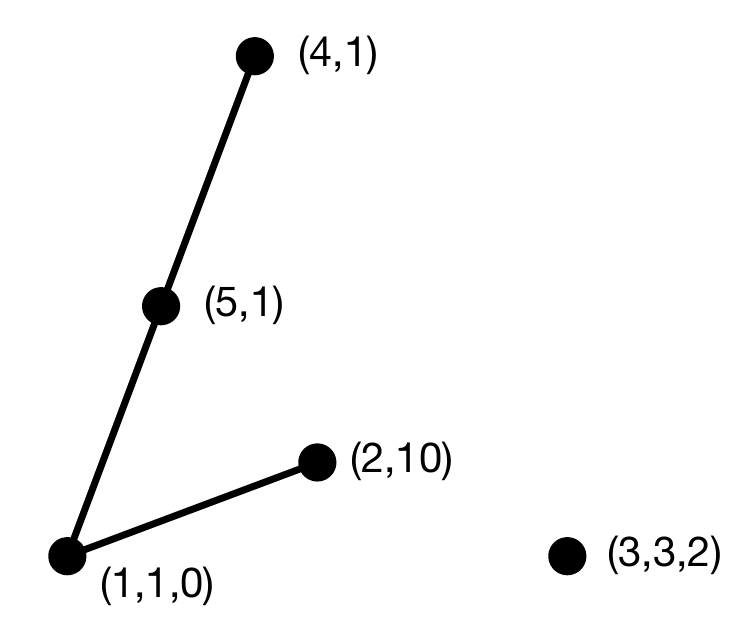}
\caption{\label{fig:forest} An example of a forest graph for the $U(5)$ theory with three flavors. The graph contains five nodes, which form two trees in this case. Each root node is labeled by $(j,\mathsf a,\mathsf k)$ while each non-root node is labeled by $(j,\mathsf k)$. In general a forest graph for the $U(5)$ theory with three flavors contains exactly five nodes and at most three trees.}
\end{figure}

 One can see that \eqref{eq:poles} defines linearly independent $N_c$ hyperplanes meeting at the unique point in $\mathbb C^{N_c}$. That intersection point can be represented by a labeled forest graph of $N_c$ nodes, which is a graph of multiple trees whose nodes are labeled by the numbers from 1 to $N_c$. If $z_j$ takes the first type of a pole, the $j$-th node will be the root node of a tree, denoted by $(j,\mathsf a,\mathsf k)$ at Figure 1 with $\mathsf a$ taking the value from $1,\ldots,N_f$ and $\mathsf k$ being a nonnegative integer. We denote $\mathsf a$ and $\mathsf k$ assigned to the $j$-th node by $a_j$ and $k_j$. If $z_j$ takes the second type of a pole, the $j$-th node will be a child node of the $i$-th node, denoted by $(j,\mathsf k)$ at Figure 1.
For instance, let us consider a $U(5)$ theory with three flavors. The forest graph corresponding to a pole determined by
\begin{align}
\begin{aligned}
z_1 &= t_1^{-1} \tau^{-1} x^{-|m_1|}, \\
z_2 &= z_1 \upsilon^{-1} x^{-|m_2-m_1|-20}, \\
z_3 &= t_3^{-1} \tau^{-1} x^{-|m_3|-4}, \\
z_4 &= z_5 \upsilon^{-1} x^{-|m_5-m_4|-2}, \\
z_5 &= z_1 \upsilon^{-1} x^{-|m_4-m_1|-2}
\end{aligned}
\end{align}
is shown in figure \ref{fig:forest}.
In general a forest graph for the $U(5)$ theory with three flavors contains exactly five nodes and at most three trees, which corresponds to three flavors.

Now let us define a map $\mathfrak p$ such that $\mathfrak p(j) = i$ if the $j$-th node is a child node of the $i$-th node. For convenience, we also assign $\mathfrak p(j) = 0$ when the $j$-th node is a root node. We also define $\mathfrak p^n= \mathfrak p\circ \mathfrak p\circ \cdots $, thus acting $\mathfrak p$ $n$ times. If we act $\mathfrak p$ iteratively, the initial position of the node is descending toward the tree node it belongs in Figure 1. $\mathfrak p(4) = 5, \mathfrak p^2(4) = 1$ and so on. Then the intersection point can be written as
\begin{align} \label{eq:intersection}
z_j = t_{b_j}^{-1} \tau^{-1} \upsilon^{-l_j+1} x^{-\sum_{n = 0}^{l_j-1}(|m_{\mathfrak p^n(j)}-m_{\mathfrak p^{n+1}(j)}|+2 k_{\mathfrak p^n(j)})}
\end{align}
where $b_j=a_i$ when $j$-th child  node is connected to the $i$-th root node. Formally one can define $b_j = a_{\mathfrak p^{l_j-1}(j)}$ and $l_j$ is the level of the $j$-th node, which is the smallest positive integer such that $\mathfrak p^{l_j}(j) = 0$. For example, a root node has the level 1. Note that the nodes in the same tree have the same $b_j$. Also we have defined $m_0 = 0$. One should note that the above poles do not guarantee the non-vanishing residues. Indeed, we will see that the non-vanishing residues come from the forest graphs with just one-branch trees.\footnote{We call a tree a one-branch tree if each node has at most one child node.}

Now we need to evaluate the residue at each intersection point. We slightly modify the expressions of the 1-loop contributions, which makes evaluating the residue easier. Firstly one can align the monopole charges as $m_1 \geq \ldots \geq m_{N_c}$ using the Weyl symmetry of the gauge group. Then the 1-loop contributions of the vector multiplet and the adjoint chiral multiplet can be written as follows:
\begin{gather}
\begin{aligned} \label{eq:vector}
\quad Z_\text{vector}(x,z,m) &= \prod_{\substack{i,j = 1 \\ (i\neq j)}}^{N_c} x^{-|m_i-m_j|/2} \left(1-z_i z_j^{-1} x^{|m_i-m_j|}\right) \\
&= \prod_{i < j}^{N_c} x^{-(m_i-m_j)} \left(1-z_i z_j^{-1} x^{m_i-m_j}\right) \left(1-z_j z_i^{-1} x^{m_i-m_j}\right),
\end{aligned} \\
\begin{aligned} \label{eq:adjoint}
&\quad Z_{X}(x,\upsilon,z,m) \\
&= \prod_{i,j = 1}^{N_c} \left(x^{-1} \upsilon\right)^{-|m_i-m_j|/2} \frac{\left(z_i^{-1} z_j \upsilon^{-1} x^{2+|m_i-m_j|};x^2\right)_\infty}{\left(z_i z_j^{-1} \upsilon x^{|m_i-m_j|};x^2\right)_\infty} \\
&= \left(\frac{\left(\upsilon^{-1} x^2;x^2\right)_\infty}{\left(\upsilon;x^2\right)_\infty}\right)^{N_c} \prod_{i < j}^{N_c} \left(x^{-1} \upsilon\right)^{-(m_i-m_j)} \frac{\left(z_i^{-1} z_j \upsilon^{-1} x^{2+m_i-m_j};x^2\right)_\infty}{\left(z_i z_j^{-1} \upsilon x^{m_i-m_j};x^2\right)_\infty} \frac{\left(z_j^{-1} z_i \upsilon^{-1} x^{2+m_i-m_j};x^2\right)_\infty}{\left(z_j z_i^{-1} \upsilon x^{m_i-m_j};x^2\right)_\infty}.
\end{aligned}
\end{gather}
Furthermore, using the identity
\begin{align}
\prod_{j = 1}^{N_c} \left(x^{-1} (-z_j) t_a \tau\right)^{(|m_j|-m_j)/2} \frac{\left(z_j^{-1} t_a^{-1} \tau^{-1} x^{2+m_j};x^2\right)_{(|m_j|-m_j)/2}}{\left(z_j t_a \tau x^{m_j};x^2\right)_{(|m_j|-m_j)/2}} = 1,
\end{align}
the contributions of the fundamental and antifundamental chiral multiplets are written as follows:
\begin{align}
&\begin{aligned} \label{eq:fund}
Z_{Q_a}(x,t,\tau,z,m) &= \prod_{j = 1}^{N_c} \left(x^{-1} z_j t_a \tau\right)^{-|m_j|/2} \frac{\left(z_j^{-1} t_a^{-1} \tau^{-1} x^{2+|m_j|};x^2\right)_\infty}{\left(z_j t_a \tau x^{|m_j|};x^2\right)_\infty} \\
&= \prod_{j = 1}^{N_c} \left(x^{-1} (-z_j) t_a \tau\right)^{-m_j/2} \frac{\left(z_j^{-1} t_a^{-1} \tau^{-1} x^{2+m_j};x^2\right)_\infty}{\left(z_j t_a \tau x^{m_j};x^2\right)_\infty},
\end{aligned} \\
&\begin{aligned} \label{eq:anti}
Z_{\tilde Q^{\tilde b}}(x,\tilde t,\tau,z,m) &= \prod_{j = 1}^{N_c} \left(x^{-1} z_j^{-1} \tilde t_a \tau\right)^{-|m_j|/2} \frac{\left(z_j \tilde t_a^{-1} \tau^{-1} x^{2+|m_j|};x^2\right)_\infty}{\left(z_j^{-1} \tilde t_a \tau x^{|m_j|};x^2\right)_\infty} \\
&= \prod_{j = 1}^{N_c} \left(x^{-1} (-z_j)^{-1} \tilde t_a \tau\right)^{-m_j/2} \frac{\left(z_j \tilde t_a^{-1} \tau^{-1} x^{2+m_j};x^2\right)_\infty}{\left(z_j^{-1} \tilde t_a \tau x^{m_j};x^2\right)_\infty}.
\end{aligned}
\end{align}
We then insert \eqref{eq:intersection} into the above expressions of the 1-loop contributions and sum their product over the poles we choose, which are represented by the labeled forest graphs we explained. From now on it is convenient to use $\mathfrak m_j = m_j-m_{\mathfrak p(j)}$ instead of $m_j$. $m_j$ is written in terms of $\mathfrak m_j$ as $m_j = \sum_{n = 0}^{l_j-1}\mathfrak m_{\mathfrak p^n(j)}$. Then carefully organizing the whole expression, we observe that the index is factorized into three parts, which we call the perturbative part, the vortex part and the antivortex part respectively. The perturbative part is independent of $\mathfrak n_j = (|\mathfrak m_j|+\mathfrak m_j)/2+k_j$ and $\bar{\mathfrak n}_j = (|\mathfrak m_j|-\mathfrak m_j)/2+k_j$. The vortex part depends on $\mathfrak n_j$ but not on $\bar{\mathfrak n}_j$ while the antivortex part depends on $\bar{\mathfrak n}_j$ but not on $\mathfrak n_j$.  The detailed computation for each part is relegated to appendix \ref{sec:detailed computation}. After the computation, the superconformal index is written in the following factorized form:
\begin{align}
I(x,t,\tilde t,\tau,\upsilon,w) = \sum_{\substack{p_a \geq 0,\\\sum_a p_a = N_c}} I_\text{pert}^{(p_a)}(x,t,\tilde t,\tau,\upsilon) Z_\text{vortext}^{(p_a)}(x,t,\tilde t,\tau,\upsilon,\mathfrak w) Z_\text{antivortex}^{(p_a)}(x,t,\tilde t,\tau,\upsilon,\mathfrak w) \label{facteq}
\end{align}
where $\mathfrak w = (-1)^{-\kappa-\frac{N_f-N_a}{2}} w$ and $p_a$'s satisfy $\sum_{a = 1}^{N_f} p_a = N_c$. The perturbative part is given by
\begin{align}
\begin{aligned}
&\quad I_\text{pert}^{(p_a)}(x,t = e^{i M},\tilde t,\tau,\upsilon = e^{i \nu}) \\
&= \left(\prod_{a,b = 1}^{N_f} \prod_{q = 1}^{p_a} \prod_{\substack{r = 1 \\ (\neq q \text{ if } a = b)}}^{p_b} 2 \sinh \frac{i M_{a}-i M_{b}+i \nu (q-r)}{2}\right) \left(\prod_{a,b = 1}^{N_f} \prod_{q = 1}^{p_a} \prod_{r = 1}^{p_b} \frac{\left(t_{a} t_{b}^{-1} \upsilon^{q-r-1} x^2;x^2\right)_\infty}{\left(t_{a}^{-1} t_{b} \upsilon^{-q+r+1};x^2\right)'_\infty}\right) \\
&\quad \times \left(\prod_{a = 1}^{N_f} \prod_{q = 1}^{p_a} \frac{\prod_{b = 1}^{N_f} \left(t_a t_b^{-1} \upsilon^{q-1} x^2;x^2\right)_\infty}{\prod_{b = 1}^{N_a} \left(t_a \tilde t_b \tau^2 \upsilon^{q-1};x^2\right)_\infty} \frac{\prod_{b = 1}^{N_a} \left(t_a^{-1} \tilde t_b^{-1} \tau^{-2} \upsilon^{-q+1} x^2;x^2\right)_\infty}{\prod_{b = 1}^{N_f} \left(t_a^{-1} t_b \upsilon^{-q+1};x^2\right)'_\infty}\right)
\end{aligned}
\end{align}
where the prime symbol indicates that the zero factor in the $q$-Pochhammer symbol is omitted. Next the vortex part and the antivortex part are given by
\begin{align}
Z_\text{vortex}^{(p_a)}(x,t,\tilde t,\tau,\upsilon,\mathfrak w) &= \sum_{\mathfrak n_j \geq 0} \mathfrak w^{\sum_{j = 1}^{N_c} \sum_{n = 0}^{l_j-1} \mathfrak n_j^n} \mathfrak Z^{(p_a)}_{(\mathfrak n_j)}(x,t,\tilde t,\tau,\upsilon), \\
Z_\text{antivortex}^{(p_a)}(x,t,\tilde t,\tau,\upsilon,\mathfrak w) &= \sum_{\bar{\mathfrak n}_j \geq 0} \mathfrak w^{-\sum_{j = 1}^{N_c} \sum_{n = 0}^{l_j-1} \bar{\mathfrak n}_j^n} \mathfrak Z^{(p_a)}_{(\bar{\mathfrak n}_j)}(x^{-1},t^{-1},\tilde t^{-1},\tau^{-1},\upsilon^{-1}),
\end{align}
\begin{align}
\begin{aligned}
&\quad \mathfrak Z^{(p_a)}_{(\mathfrak n_j)}(x = e^{-\gamma},t = e^{i M},\tilde t = e^{i \tilde M},\tau = e^{i \mu},\upsilon = e^{i \nu}) \\
&= e^{-S^{(p_a)}_{(\mathfrak n_j)}(x,t,\tau,\upsilon)} \left(\prod_{a,b = 1}^{N_f} \prod_{q = 1}^{p_a} \prod_{\substack{r = 1 \\ (\neq q \text{ if } a = b)}}^{p_b} \prod_{k = 1}^{\sum_{n = 1}^r \mathfrak n_{(b,n)}} \frac{\sinh \frac{i M_a-i M_b+i \nu (q-r)+2 \gamma k}{2}}{\sinh \frac{i M_a-i M_b+i \nu (q-r)+ 2 \gamma (k-1-\sum_{n = 1}^q \mathfrak n_{(a,n)})}{2}}\right) \\
&\quad \times \left(\prod_{a,b = 1}^{N_f} \prod_{q = 1}^{p_a} \prod_{\substack{r = 1 \\ (\neq q \text{ if } a = b)}}^{p_b} \prod_{k = 1}^{\sum_{n = 1}^r \mathfrak n_{(b,n)}} \frac{\sinh \frac{i M_a-i M_b+i \nu (q-r-1)+2 \gamma (k-1-\sum_{n = 1}^q \mathfrak n_{(a,n)})}{2}}{\sinh \frac{i M_a-i M_b+i \nu (q-r+1)+2 \gamma k}{2}}\right) \\
&\quad \times \left(\prod_{b = 1}^{N_f} \prod_{r = 1}^{p_b} \prod_{k = 1}^{\sum_{n = 1}^r \mathfrak n_{(b,n)}} \frac{\prod_{a = 1}^{N_a} \sinh \frac{-i \tilde M_a-i M_b-2 i \mu-i \nu (r-1)+2 \gamma (k-1)}{2}}{\prod_{a = 1}^{N_f} \sinh \frac{i M_a-i M_b-i \nu (r-1)+2 \gamma k}{2}}\right),
\end{aligned}
\end{align}
where
\begin{align}
e^{-S^{(p_a)}_{(\mathfrak n_j)}(x,t,\tau,\upsilon)} = \prod_{b = 1}^{N_f} \prod_{r = 1}^{p_b} \left(t_b \tau \upsilon^{r-1} x^{\sum_{n = 1}^r \mathfrak n_{(b,n)}}\right)^{\kappa \sum_{n = 1}^r \mathfrak n_{(b,n)}}
\end{align}
$\mathfrak n_{(a,q)}$ is a shorthand notation for $\mathfrak n_{q+\sum_{b = 1}^{a-1} p_b}$. Note that the antivortex part is obtained from the vortex part by inverting all the fugacities, $x,t,\tilde t,\tau,\upsilon,w \rightarrow x^{-1},t^{-1},\tilde t^{-1},\tau^{-1},\upsilon^{-1},\mathfrak w^{-1}$. We expect that this is the vortex partition function on $\mathbb R^2 \times S^1$ of the $\mathcal N = 2$ $U(N_c)_\kappa$ gauge theory with $N_f$ fundamental, $N_a$ antifundamental and one adjoint matter under the condition $|\kappa| \leq \frac{N_f-N_a}{2}$. As far as we know, there is no explicit computation of the vortex partition function of a 3d $\mathcal N = 2$ theory with an adjoint matter. Nevertheless, from the Higgs branch localization of the 3d superconformal index \cite{Benini:2013yva,Fujitsuka:2013fga} or the consideration of holomorphic blocks \cite{Beem:2012mb},  what
we call the ``vortex'' (``antivortex'') parts should correspond to the vortex (antivortex) partition function of a 3d $\mathcal N = 2$ theory with an adjoint matter. For the special case of $\mathcal N = 4$ theories, the above reduces the known results of the vortex partition function \cite{Kim:2012uz}.

\section{3d $\mathcal N = 4$ Seiberg-like duality} \label{sec:N=4}
Our first application is the 3d $\mathcal N = 4$ $U(N_c)$ gauge theories with fundamental matters. The theory has the $SO(4)_R$ $R$-symmetry as well as the global symmetry $SU(N_f) \times U(1)_T$. Those theories are classified into 3 classes according to the number of flavors: good, ugly, bad \cite{Gaiotto:2008ak,Kapustin:2010mh}. A theory is ``good'' if the number of flavors $N_f$  and the rank of the gauge group $N_c$ satisfy $ N_f > 2 N_c-1$. In that case the monopole operators appearing in the theory have the $R$-charges larger than 1/2. Accordingly they have the conformal dimensions  larger than 1/2 in IR, which is required for the unitarity of an IR fixed point. A theory is ``ugly'' if $N_f = 2 N_c-1$. In that case the theory has a  monopole operator having the $R$-charges, hence the conformal dimensions, 1/2. Therefore, they are decoupled in the IR fixed point. It is also known that the theory has a Seiberg-like dual description, the $U(N_c-1)$ theory with $N_f$ fundamental matters and one decoupled free twisted hypermultiplet \cite{Gaiotto:2008ak,Kapustin:2010mh}. The original and dual theory flow to the same IR fixed point. The decoupled monopole operator of the original theory corresponds to the decoupled hypermultiplet of the dual theory. A theory is ``bad'' if $N_c \leq N_f < 2 N_c-1$. Such a theory has monopole operators whose UV $R$-charges are less than 1/2, and even less than zero. If those UV $R$-charges are preserved to IR along the RG flow, the corresponding monopole operators have the unitarity violating conformal dimensions. However, if there is
accidental global symmetries in IR, the R symmetry in IR can be mixed with such global symmetries. Hence the IR $R$-charge i.e., the conformal dimension of the monopole operators does not need to be less than 1/2. Indeed, those monopole operators are expected to have the conformal dimensions exactly 1/2 and thus decouple from the theory. Therefore, a ``bad'' theory has the IR fixed point consisting of a decoupled free sector and an interacting sector. Interestingly there is another UV description describing those two sectors separately. In that description, the decoupled sector is described by $2 N_c-N_f$ free twisted hypermultiplets while the interacting sector is described by the $U(N_f-N_c)$ theory with $N_f$ fundamental matters \cite{Kim:2012uz,Yaakov:2013fza,Gaiotto:2013bwa}. Note that there is no interacting sector if $N_f = N_c$. This is the $\mathcal N = 4$ version of the conjectured 3d Seiberg-like duality for $U(N_c)$ gauge theories.
\begin{table}[tbp]
\centering
\begin{tabular}{|c|cccc|}
\hline
 & $U(1)_R$ & $SU(N_f)$ & $U(1)_A$ & $U(1)_T$ \\
\hline
$Q$ & $1/2$ & $\overline{\mathbf N_f}$ & 1 & 0 \\
$\tilde Q$ & $1/2$ & $\mathbf N_f$ & 1 & 0 \\
$X$ & $1$ & $\mathbf 1$ & -2 & 0 \\
\hline
$V_{i,\pm}$ & $\frac{1}{2} N_f-N_c+1+i$ & $\mathbf 1$ & $-N_f+2 N_c-2-2 i$ & $\pm1$ \\
\hline
\end{tabular}
\caption{\label{tab:N=4 charges} The global symmetry charges of chiral operators in the $\mathcal N = 2$ language. The $U(1)_R$ and $U(1)_A$ charges are given by $R_H+R_C$ and $2 R_H-2 R_c$ respectively where $R_H$ and $R_C$ are spins of the $SO(4)_R = SU(2)_H \times SU(2)_C$ $R$-symmetry.}
\end{table}

Although the localization computations of various partition functions usually provide powerful tools for testing dualities, they have convergence issues for the ``bad'' theories. The definition \eqref{eq:index def} defines the index as a power series in $x$. However, the matrix integral \eqref{eq:matrix integral} is not analytic at $x = 0$ for bad theories since it contains the BPS monopole operators of negative conformal dimensions. In order to avoid those issues, one can try to find a quantity that is free of the convergence issue, e.g., the vortex partition function on $\mathbb R^2 \times S^1$ in \cite{Kim:2012uz}, or to use the analytically continued version of the partition function, e.g., the $S^3_b$ partition function in \cite{Yaakov:2013fza}. Our strategy is similar.  In fact, since the factorized index is written in terms of the vortex partition functions, the comparison of the indices of a duality pair reduces to that of the vortex partition functions, which is numerically performed in \cite{Kim:2012uz}. Here we provide an analytic proof of the agreement of the vortex partition functions and that of the perturbative parts as well. Thus, we provide a complete proof of the index agreement. This also clarifies  the role of accidental $R$-symmetry relating
 UV and IR quantities. Previously  comparisons have been made for some limits of the superconformal index, which correspond to the Hilbert series \cite{Cremonesi:2013lqa,Razamat:2014pta}.

\subsection{SCI under duality}
The superconformal index of a 3d $\mathcal N = 4$ theory can be defined as
\begin{align}
I\left(x,t_i\right) = \mathrm{Tr} (-1)^F \exp(-\beta'\{Q,S\}) x^{E+j} y^{R_H-R_C} \left(\prod_i t_i^{F_i}\right)
\end{align}
where $R_H$ and $R_C$ are the charges of the Cartans of $SO(4)_R = SU(2)_H \times SU(2)_C$ $R$-symmetry and the other variables have the same meaning as in the $\mathcal N= 2$ case. The BPS condition is given by
\begin{align}
\{Q,S\} = E-R_H-R_C-j \geq 0.
\end{align}
It is convenient to set $y = \tau^2$ such that $\tau$ plays the role of the $U(1)_A$ fugacity in the $\mathcal N = 2$ language. The factorized index of a $\mathcal N = 4$ theory can be obtained from the $\mathcal N = 2$ result by appropriately adjusting fugacities. Firstly we substitute $\tilde t_a = t_a^{-1}$ and shift $\tau \rightarrow \tau x^{1/2}$ because the fundamental and antifundamental chiral multiplets form the hypermultiplets whose $U(1)_R$ charges are fixed to 1/2. In addition, since the adjoint chiral multiplet is now a part of the $\mathcal N = 4$ vector multiplet, which is a triplet of $SU(2)_C$, it has $U(1)_R$ charge 1 and the $U(1)_A$ charge $-2$. There is no $U(1)_X$ symmetry independently rotating the adjoint chiral multiplet because of the superpotential term $\tilde Q X Q$. Therefore, we need to substitute $\upsilon = \tau^{-2} x$. Then it is easy to see that $I_\text{pert}$ vanishes if $p_a > 1$ in eq. (\ref{facteq}). Thus, for a $\mathcal N = 4$ theory the nontrivial poles are labeled by the choices of $N_c$ distinct numbers between 1 and $N_f$, which is the same as in the $\mathcal N = 2$ case without an adjoint matter \cite{Hwang:2012jh}.

Therefore, the superconformal index of the 3d $\mathcal N = 4$ $U(N_c)$ gauge theory with $N_f$ fundamental matters can be written as follows:
\begin{align}
I(x,t,\tau,w)=\sum_{\substack{1\leq b_1<\ldots \\<b_{N}\leq N_f}} I_\text{pert}^{\{b_j\}}\left(x,t\right) Z_\text{vortex}^{\{b_j\}}\left(x,t,\tau,w\right) Z_\text{antivortex}^{\{b_j\}}\left(x,t,\tau,w\right)
\end{align}
where
\begin{align}
\begin{aligned}
&\quad I_\text{pert}^{\{b_j\}}(x,t=e^{i M},\tau) \\
&= \left(\prod_{\substack{i,j = 1 \\ (i\neq j)}}^{N_c} 2 \sinh \frac{i M_{b_i}-i M_{b_j}}{2}\right) \left(\prod_{j = 1}^{N_c} \frac{\prod_{a = 1 (\neq b_j)}^{N_f} \left(t_{b_j} t_a^{-1} x^2;x^2\right)_\infty}{\prod_{a \in \{b_j\}^c} \left(t_{b_j} t_a^{-1} \tau^2 x;x^2\right)_\infty} \frac{\prod_{a \in \{b_j\}^c} \left(t_{b_j}^{-1} t_a \tau^{-2} x;x^2\right)_\infty}{\prod_{a = 1 (\neq b_j)}^{N_f} \left(t_{b_j}^{-1} t_a;x^2\right)_\infty}\right),
\end{aligned}
\end{align}
\begin{align}
Z_\text{vortex}^{\{b_j\}}(x,t,\tau,w) &= \sum_{n_j \geq 0} w^{\sum_{j = 1}^{N_c} n_j} \mathfrak Z_{(n_j)}^{\{b_j\}}(x,t,\tau), \\
Z_\text{antivortex}^{\{b_j\}}(x,t,\tau,w) &= \sum_{n_j \geq 0} w^{-\sum_{j = 1}^{N_c} n_j} \mathfrak Z_{(n_j)}^{\{b_j\}}(x,t,\tau),
\end{align}
\begin{align}
\begin{aligned}
&\quad \mathfrak Z_{(n_j)}^{\{b_j\}}(x = e^{- \gamma},t = e^{i M},\tau = e^{i \mu}) \\
&= \prod_{j = 1}^{N_c} \prod_{k = 1}^{n_j} \left(\prod_{i = 1}^{N_c} \frac{\sinh \frac{i M_{b_i}-i M_{b_j}+2 i \mu+2 \gamma (k-\frac{1}{2}-n_i)}{2}}{\sinh \frac{i M_{b_i}-i M_{b_j}+2 \gamma (k-1-n_i)}{2}}\right) \left(\prod_{a \in \{b_j\}^c}^{N_f} \frac{\sinh \frac{i M_a-i M_{b_j}-2 i \mu+2 \gamma (k-\frac{1}{2})}{2}}{\sinh \frac{i M_a-i M_{b_j}+2 \gamma k}{2}}\right).
\end{aligned}
\end{align}
We will show each part is exactly the same as that of the dual theory, the $U(N_f-N_c)$ theory with $N_f$ fundamental matters
possibly with the additional hypermultiplets. Firstly let us consider the perturbative part, $I_\text{pert}^{\{b_j\}}$, which is straightforward to prove. Let us have a look at the following factor:
\begin{align*}
\prod_{j = 1}^{N_c} \prod_{a = 1 (\neq b_j)}^{N_f} \frac{\left(t_{b_j} t_a^{-1} x^2;x^2\right)_\infty}{\left(t_{b_j}^{-1} t_a;x^2\right)_\infty}.
\end{align*}
It can be decomposed into two parts as follows:
\begin{align}
\begin{aligned}
\prod_{j = 1}^{N_c} \prod_{a = 1 (\neq b_j)}^{N_f} \frac{\left(t_{b_j} t_a^{-1} x^2;x^2\right)_\infty}{\left(t_{b_j}^{-1} t_a;x^2\right)_\infty} &= \left(\prod_{\substack{i,j = 1 \\ (i \neq j)}}^{N_c} \frac{\left(t_{b_j} t_{b_i}^{-1} x^2;x^2\right)_\infty}{\left(t_{b_j}^{-1} t_{b_i};x^2\right)_\infty}\right) \left(\prod_{j = 1}^{N_c} \prod_{a \in \{b_j\}^c} \frac{\left(t_{b_j} t_a^{-1} x^2;x^2\right)_\infty}{\left(t_{b_j}^{-1} t_a;x^2\right)_\infty}\right) \\
&= \left(\prod_{\substack{i,j = 1 \\ (i \neq j)}}^{N_c} \frac{1}{1-t_{b_j}^{-1} t_{b_i}}\right) \left(\prod_{j = 1}^{N_c} \prod_{a \in \{b_j\}^c} \frac{\left(t_{b_j} t_a^{-1} x^2;x^2\right)_\infty}{\left(t_{b_j}^{-1} t_a;x^2\right)_\infty}\right).
\end{aligned}
\end{align}
Note that the flavor indices are decomposed into $N_c$ flavors and $N_f-N_c$ flavors.
Therefore, the perturbative part can be written as
\begin{align}
\begin{aligned} \label{eq:N=4 pert duality}
I_\text{pert}^{\{b_j\},N_c,N_f}(x,t=e^{i M},\tau) &= \prod_{j = 1}^{N_c} \prod_{a \in \{b_j\}^c} \frac{\left(t_{b_j} t_a^{-1} x^2;x^2\right)_\infty}{\left(t_{b_j} t_a^{-1} \tau^2 x;x^2\right)_\infty} \frac{\left(t_{b_j}^{-1} t_a \tau^{-2} x;x^2\right)_\infty}{\left(t_{b_j}^{-1} t_a;x^2\right)_\infty} \\
&= \prod_{a \in \left\{\tilde b_j\right\}^c} \prod_{j = 1}^{N_f-N_c} \frac{\left(t_a t_{\tilde b_j}^{-1} x^2;x^2\right)_\infty}{\left(t_a t_{\tilde b_j}^{-1} \tau^2 x;x^2\right)_\infty} \frac{\left(t_a^{-1} t_{\tilde b_j} \tau^{-2} x;x^2\right)_\infty}{\left(t_a^{-1} t_{\tilde b_j};x^2\right)_\infty} \\
&= I_\text{pert}^{\left\{\tilde b_j\right\},N_f-N_c,N_f}(x,t^{-1}=e^{-i M},\tau)
\end{aligned}
\end{align}
where $\{\tilde b_j\}$ is defined by $\{\tilde b_j\} = \{b_j\}^c$. This shows that the perturbative part with a given choice of $\{b_j\}$ is the same as that of the dual theory with the conjugate choice $\{\tilde b_j\} = \{b_j\}^c$.

Now let us examine the vortex part. As a first step, we show the following identity:
\begin{align} \label{eq:N=4 identity}
Z_\text{vortex}^{\{b_j\},N_c,2 N_c}(x,t,\tau,w) = Z_\text{vortex}^{\{b_j\}^c,N_c,2 N_c}(x,t^{-1},\tau,w).
\end{align}
The left hand side is written as
\begin{align*}
\sum_{n \geq 0} w^n \left(\sum_{\substack{n_j \geq 0,\\\sum_j n_j = n}} \mathfrak Z_{(n_j)}^{\{b_j\}}(x,t,\tau)\right)
\end{align*}
Following the method used for 2d theories \cite{Benini:2012ui}, one can show that the coefficient of $w^n$ on each side coincides with each other. Firstly one can write down the coefficient of $w^n$ as the following integral expression:
\begin{align}
\begin{aligned}
&\quad \sum_{\substack{n_j \geq 0,\\\sum_j n_j = n}} \mathfrak Z_{(n_j)}^{\{b_j\},N_c,2 N_c}(x = e^{- \gamma},t = e^{i M},\tau = e^{i \mu}) \\
&= \sum_{\substack{n_j \geq 0,\\\sum_j n_j = n}} \prod_{j = 1}^{N_c} \prod_{k = 1}^{n_j} \left(\prod_{i = 1}^{N_c} \frac{\sinh \frac{i M_{b_i}-i M_{b_j}+2 i \mu+2 \gamma (k-\frac{1}{2}-n_i)}{2}}{\sinh \frac{i M_{b_i}-i M_{b_j}+2 \gamma (k-1-n_i)}{2}}\right) \left(\prod_{a \in \{b_j\}^c} \frac{\sinh \frac{i M_a-i M_{b_j}-2 i \mu+2 \gamma (k-\frac{1}{2})}{2}}{\sinh \frac{i M_a-i M_{b_j}+2 \gamma k}{2}}\right) \\
&= \frac{1}{n!} \int_{\phi_I = 0}^{2 \pi} \left(\prod_{I = 1}^{n} \frac{d\phi_I}{2 \pi} \frac{1}{\sinh \gamma}\right) \left(\prod_{I \neq J}^{n} \frac{\sinh \frac{i \phi_I-i \phi_J}{2}}{\sinh \frac{i \phi_I-i \phi_J+2 \gamma}{2}}\right) \left(\prod_{I,J = 1}^{n} \frac{\sinh \frac{i \phi_I-i \phi_J-2 i \mu+\gamma}{2}}{\sinh \frac{i \phi_I-i \phi_J-2 i \mu-\gamma}{2}}\right) \\
&\qquad \times \prod_{I = 1}^n \frac{\left(\prod_{a \in \{b_j\}} \sinh \frac{i \phi_I-i M_a-2 i \mu}{2}\right) \left(\prod_{a \in \{b_j\}^c} \sinh \frac{i M_a-i \phi_I-2 i \mu}{2}\right)}{\left(\prod_{a \in \{b_j\}} \sinh \frac{i \phi_I-i M_a+\gamma}{2}\right) \left(\prod_{a \in \{b_j\}^c} \sinh \frac{i M_a-i \phi_I+\gamma}{2}\right)}
\end{aligned}
\end{align}
where we assume $-2 i \mu > \gamma > 0$ and $M_a$'s are real. Regarding the integration as a contour integration along the unit circle $|z_I| = |e^{i \phi_I}| = 1$, one can apply the residue theorem. Then the relevant poles inside the unit circle are
\begin{align}
i \phi_I = \left\{\begin{array}{lc}
i M_a-\gamma, & a \in \{b_j\} \\
i \phi_J-2 \gamma, & 1 \leq I (\neq J) \leq n \\
i \phi_J+2 i \mu+\gamma.  & 1 \leq I (\neq J) \leq n
\end{array} \right.
\end{align}
However, if the last type of a pole is chosen, the residue becomes zero because either $\sinh \frac{i \phi_I-i \phi_J-2 i \mu+\gamma}{2}$ or $\sinh \frac{i \phi_I-i M_a-2 i \mu}{2}$ in the numerator vanishes. Therefore, only the first two types of poles can be chosen. Then the intersection point is written as
\begin{align}
i \phi_I = i M_{b_j}-(2 k-1) \gamma, \quad k = 1,\ldots,n_j, \quad \sum_{j = 1}^{N_c} n_j = n.
\end{align}
Thus, one can check that the above integral gives rise to the original vortex partition function. On the other hand, one can also take the poles from the outside of the unit circle. In that case, the relevant poles are
\begin{align}
i \phi_I = \left\{\begin{array}{lc}
i M_a+\gamma, & a \in \{b_j\}^c \\
i \phi_J+2 \gamma, & 1 \leq I (\neq J) \leq n \\
i \phi_J-2 i \mu-\gamma,  & 1 \leq I (\neq J) \leq n
\end{array} \right.
\end{align}
but again the last type of a pole gives rise to the vanishing residue. Therefore, the nontrivial intersection point is written as
\begin{align}
i \phi_I = i M_{c_j}+(2 k-1) \gamma, \quad k = 1,\ldots,n_j, \quad \sum_{j = 1}^{N_c} n_j = n
\end{align}
where $c_j \in \{b_j\}^c$. Then one can check that the residue is given by $\sum_{\substack{n_j \geq 0,\\\sum_j n_j = n}} \mathfrak Z_{(n_j)}^{\{b_j\}^c,N_c,2 N_c}(x,t^{-1},\tau)$. Thus, we have proven the identity \eqref{eq:N=4 identity}.

In order to prove the agreement of the vortex part for $N_f < 2 N_c$, we consider the large mass behavior of the vortex partition function. Note that we have assumed $|t_a| = 1$; i.e, $M_a$ is on the real line. From now on we analytically continue $M_a$ to the whole complex plane such that its complex part corresponds to real mass of the $a$-th flavor. In appendix \ref{sec:large mass} we examine the behavior of the vortex partition function under the large real mass limit. We have two different limits of the vortex partition function depending on whether the large mass flavor is picked from $\{b_j\}$ or from $\{b_j\}^c$:
\begin{align}
&\begin{aligned} \label{eq:large mass 2}
&\quad \lim_{i M_{\mathfrak b} \rightarrow \pm\infty} Z_\text{vortex}^{\{b_j\},N_c,N_f}(x,t,\tau,w) \\
&= Z_\text{hyper}(x,\tau,w \tau^{\mp(2 N_c-N_f-1)} x^{\pm(2 N_c-N_f-1)/2}) \times Z_\text{vortex}^{\{b_j\},N_c-1,N_f-1}(x,t'',\tau,w \tau^{\pm1} x^{\mp\frac{1}{2}}),
\end{aligned} \\
&\quad \lim_{i M_{\mathfrak a} \rightarrow \pm\infty} Z_\text{vortex}^{\{b_j\},N_c,N_f}\left(x,t,\tau,w\right) = Z_\text{vortex}^{\{b_j\},N_c,N_f-1}(x,t',\tau,w \tau^{\mp1} x^{\pm1}) \label{eq:large mass 1}
\end{align}
where $Z_\text{hyper}$ is the contribution of the free twisted hypermultiplet whose square gives rise to the index of the hypermultiplet: $I_\text{hyper}(x,\tau,w) = Z_\text{hyper}(x,\tau,w) Z_\text{hyper}(x^{-1},\tau^{-1},w^{-1})$. $\mathfrak b$ and $\mathfrak a$ are chosen such that $\mathfrak b \in \{b_j\}$ and $\mathfrak a \in \{b_j\}^c$.

Using those results one can find a set of identities for $N_f < 2 N_c$. Firstly we assume $2 N_c$ is not in $\{b_j\}$. We then take the limit $i M_{2 N_c} \rightarrow \infty$ for \eqref{eq:N=4 identity}. Using \eqref{eq:large mass 1} we observe that the left hand side becomes
\begin{align}
\lim_{i M_{2 N_c} \rightarrow \infty}Z_\text{vortex}^{\{b_j\},N_c,2 N_c}(x,t,\tau,w) = Z_\text{vortex}^{\{b_j\},N_c,2 N_c-1}(x,t',\tau,w \tau^{-1} x^\frac{1}{2})
\end{align}
where $t'=(t_1,\ldots,t_{2 N_c-1})$. On the other hand, for the right hand side we should use \eqref{eq:large mass 2} because $2 N_c$ is an element of $\{b_j\}^c$. Thus, the right hand side becomes
\begin{align}
\begin{aligned}
&\quad \lim_{-i M_{2 N_c} \rightarrow -\infty}Z_\text{vortex}^{\{b_j\}^c,N_c,2 N_c}(x,t^{-1},\tau,w) \\
&= Z_\text{hyper}(x,\tau,w \tau^{-1} x^\frac{1}{2}) \times Z_\text{vortex}^{\{b_j\}^c,N_c-1,2 N_c-1}(x,t'^{-1},\tau,w \tau^{-1} x^\frac{1}{2})
\end{aligned}
\end{align}
where the additional contribution of one free twisted hypermultiplet appears. Repeating this procedure, we eventually obtain the following identity:
\begin{align}
\begin{aligned}
&\quad Z_\text{vortex}^{\{b_j\},N_c,N_f}(x,t,\tau,w \tau^{-(2 N_c-N_f)} x^{(2 N_c-N_f)/2}) \\
&= Z_\text{vortex}^{\{b_j\}^c,N_f-N_c,N_f}(x,t^{-1},\tau,w \tau^{-(2 N_c-N_f)} x^{(2 N_c-N_f)/2}) \times \prod_{i = 1}^{2 N_c-N_f}Z_\text{hyper}(x,\tau,w \tau^{-(2 i-1)} x^{(2 i-1)/2}),
\end{aligned}
\end{align}
or equivalently,
\begin{align}
\begin{aligned}\label{eq:N=4 vpf duality}
&\quad Z_\text{vortex}^{\{b_j\},N_c,N_f}(x,t,\tau,w) \\
&= Z_\text{vortex}^{\{b_j\}^c,N_f-N_c,N_f}(x,t^{-1},\tau,w) \times \prod_{i = 1}^{2 N_c-N_f}Z_\text{hyper}(x,\tau,w \tau^{2 N_c-N_f-2 i+1} x^{-(2 N_c-N_f-2 i+1)/2}).
\end{aligned}
\end{align}
Note that the procedure terminates at $N_f=N_c$ because if $N_f = N_c$, $t$ dependency of the identity completely disappears. Thus, the above identity holds for $N_c \leq N_f \leq 2 N_c$.

From \eqref{eq:N=4 pert duality} and \eqref{eq:N=4 vpf duality}, we conclude that the superconformal index of the $\mathcal N = 4$ $U(N_c)$ gauge theory with $N_c \leq N_f \leq 2 N_c$ flavors satisfies the following identity:
\begin{align}
\begin{aligned}\label{eq:index duality}
&\quad I^{N_c,N_f}(x,t,\tau,w) \\
&= I^{N_f-N_c,N_f}(x,t^{-1},\tau,w) \times \prod_{i = 1}^{2 N_c-N_f}I_\text{hyper}(x,\tau,w \tau^{2 N_c-N_f-2 i+1} x^{-(2 N_c-N_f-2 i+1)/2}),
\end{aligned}
\end{align}
which is strong evidence of the Seiberg-like duality for 3d $\mathcal N = 4$ theories. Note that a similar identity is observed for the $S^3_b$ partition functions \cite{Yaakov:2013fza}. In addition, one can rewrite the free twisted hypermultiplet part as follows:
\begin{align}
\begin{aligned} \label{eq:free hypers}
&\prod_{i = 1}^{2 N_c-N_f}I_\text{hyper}(x,\tau,w \tau^{2 N_c-N_f-2 i+1} x^{-(2 N_c-N_f-2 i+1)/2}) \\
&= \prod_{i = 1}^{2 N_c-N_f}I_\text{hyper}(x,\tau^{-2 N_c+N_f+2 i} x^{(2 N_c-N_f-2 i+1)/2},w).
\end{aligned}
\end{align}
Note that if we assign generic $U(1)_A$ and $U(1)_R$ charges $A$ and $R$ to the twisted hypermultiplet, its index is given by
\begin{align}
I_\text{hyper}(x,\tau^{-A} x^{-R+\frac{1}{2}},w) = \mathrm{PE}\left[\frac{\tau^A w x^R-\tau^{-A} w^{-1} x^{2-R}}{1-x^2}\right] \times \mathrm{PE}\left[w \leftrightarrow w^{-1}\right].
\end{align}
Thus, \eqref{eq:free hypers} is the index contribution of free twisted hypermultiplets of the $U(1)_A$ charges $2 N_c-N_f-2 i$ and the $U(1)_R$ charges $-(2 N_c-N_f-2 i)/2$ with $i = 1,\ldots,2 N_c-N_f$. Comparing with table \ref{tab:N=4 charges}, those charges are exactly the $U(1)_A$ and $U(1)_R$ charges of monopole operators $V_{i-1,\pm}$ in the original theory. Note that the free hypermultiplets
carry nonstandard $U(1)_A$ and $R$-charges. Such features are also present in other dualities with accidental symmetries in IR \cite{Kapustin:2011vz}. The peculiar feature of the current example is that some of the operators are carrying negative $R$-charges in UV. Thus, the above computation  supports the expectation that monopole operators of negative UV $R$-charges decouple  IR.

\subsection{Examples}
\subsubsection{$N_c = N_f$}
Let us examine some examples. Firstly we consider the $U(1)$ theory with one flavor. This theory is ``ugly'', and its monopole operator has the conformal dimension 1/2. Its Seiberg-like dual theory is a free theory of one twisted hypermultiplet. From \eqref{eq:index duality} one can see that their indices are exactly the same:
\begin{align}
I^{1,1}(x,1,\tau,w) &= I_\text{hyper}(x,\tau,w).
\end{align}
Let us have a look at the chiral ring elements, especially the generators, which describe the moduli space of the theory. Candidates of the chiral ring generators of the original theory are the meson operator $M = Q \tilde Q$, two monopole operator $V_{0,\pm}$ and the Casimir invariant $\mathrm{tr} X$. However, the meson operator is lifted due to the superpotential term $\tilde Q X Q$ and there is a nontrivial relation between the monopole operators and the Casimir invariant: $V_{0,+} V_{0,-} = \mathrm{tr} X$ \cite{Cremonesi:2013lqa}. Therefore, the only chiral ring generators are the monopole operators. Under the duality those monopole operators are mapped to the two chiral multiplets in the free twisted hypermultiplet, which are again two chiral ring generators of the dual theory. Their contribution appears in the index as the lowest nontrivial energy term:
\begin{align}
I^{1,1}(x,1,\tau,w) = 1+\sqrt{x} \left(\frac{w}{\tau }+\frac{1}{\tau  w}\right)+x
   \left(\frac{1}{\tau ^2}+\frac{w^2}{\tau ^2}+\frac{1}{\tau ^2
   w^2}\right)+O\left(x^{3/2}\right)
\end{align}
In addition, it is known that for a superconformal theory with large enough supersymmetry, the global symmetry as well as the $R$-symmetry can be read off from the superconformal index because the conserved currents form supermultiplets \cite{Bashkirov:2010kz,Bashkirov:2011fr,Cheon:2012be}. For a $\mathcal N = 4$ theory, the global symmetry current forms a supermultiplet whose lowest component is a spacetime scalar in the representation $\mathbf{3} \times \text{adj}$ of $SO(4)_R \times G$ where $G$ is the global symmetry of the theory. It is decomposed under $U(1)_R \times U(1)_A \times G \subset SO(4)_R \times G$ as $\text{adj}_{1,-1}+\text{adj}_{0,0}+\text{adj}_{-1,1}$ where adj means the adjoint representation of $G$ while
the subscripts denote the charges of $U(1)_R$ and $U(1)_A$. The index captures the BPS sector of that, $\text{adj}_{1,-1}$, whose index contribution is therefore $\tau^{-2} \chi^G_\text{adj} x$. In the current example, $x^1$ term in the superconformal index is $x \tau^{-2} (w^2+1+w^{-2}) = x \tau^{-2} \chi^{SU(2)}_\text{adj}$, which forms the character of the $SU(2)$ adjoint representation. It tells us that the $U(1)_T$ symmetry of the original theory is enhanced to $SU(2)$.

Another example is the $U(2)$ theory with two flavors whose dual theory is the free theory of two twisted hypermultiplets. This theory has two pairs of monopole operators of the UV $R$-charges 0 and 1: $V_{0,\pm},V_{1,\pm}$. It is the simplest example of a ``bad'' theory, for which there exist monopole operators of unitarity violating UV $R$-charges. Thus, if we compute its index using the UV content, we have infinitely many zero energy terms:
\begin{align}\label{eq:infinite vacuua}
I^{2,2}(x,t,\tau,w) = \sum_{n \geq 0} \sum_{\bar n \geq} w^{n-\bar n}+O(x)
\end{align}
where the term $w^{n-\bar n}$ corresponds to the operator $V_{0,+}^n V_{0,-}^{\bar n}$. For this reason, usual perturbative analysis of the index by series expansion is not allowed in this case. Indeed, the index we compute is not fully refined because there should be additional IR symmetries which are not visible in UV. Nevertheless, since we have the analytic identity \eqref{eq:index duality}. We can observe that the indices of the original and dual theories coincide if we assign specific global charges to the free twisted hypermultiplets on the dual side:
\begin{align}\label{eq:2,2}
\quad I^{2,2}(x,t,\tau,w) = I_\text{hyper}(x,\tau,w \tau x^{-\frac{1}{2}}) \times I_\text{hyper}(x,\tau,w \tau^{-1} x^\frac{1}{2}).
\end{align}
Furthermore, this identity allows us to compare IR symmetry and UV symmetry of the original theory. First of all, since the dual theory is free, its index is completely determined by the UV content although the right hand side of \eqref{eq:2,2} is not fully refined. We can introduce fugacities $w_1$ and $w_2$ for $U(1)_{B,1}$ and $U(1)_{B,2}$, each of which rotates each hypermultiplet independently. Then the right hand side of \eqref{eq:2,2} can be refined as
\begin{align} \label{eq:free2}
I_\text{hyper}(x,\tau,w_1) \times I_\text{hyper}(x,\tau,w_2).
\end{align}
Due to duality the refined index \eqref{eq:free2} should be that of the original theory.\footnote{The refined index is the index with the fugacities of all the global symmetry.} The index \eqref{eq:free2} is reduced to the unrefined index \eqref{eq:2,2} by $w_1 \rightarrow w \tau x^{-\frac{1}{2}},w_2 \rightarrow w \tau^{-1} x^\frac{1}{2}$. Therefore, one can identify the UV $R$-charges and global charges with the IR $R$-charges and global charges as follows:
\begin{align}
R^{UV} &= R^{IR}-\frac{1}{2} B_1^{IR}+\frac{1}{2} B_2^{IR}, \\
A^{UV} &= A^{IR}+B_1^{IR}-B_2^{IR}, \\
T^{UV} &= B_1^{IR}+B_2^{IR}
\end{align}
where $R^{UV/IR}$ and $A^{UV/IR}$ are $U(1)_R \times U(1)_A$ charges in UV/IR; $T^{UV}$ is $U(1)_T$ charge in UV while $B_1^{IR},B_2^{IR}$ are $U(1)_{B,1} \times U(1)_{B,2}$ charges in IR. Note that only the diagonal $U(1)$ of $U(1)_{B,1} \times U(1)_{B,2}$, which corresponds to $U(1)_T$, is visible in UV on the original side. From the above equations, with the standard $R$-charge and $U(1)_A$ assignment of free hypermultiplets, one can read off the original UV charges. For example, one  finds that R-charges of two hypermultiplets are $0,1$, recovering the previous assignments. In addition, we can again read off the global symmetry from the superconformal index. Series expanding \eqref{eq:free2}:
\begin{align}
\begin{aligned}\label{eq:two free index}
&\quad I_\text{hyper}(x,\tau,w_1) \times I_\text{hyper}(x,\tau,w_2) \\
&= 1+\sqrt{x} \left(\frac{w_1}{\tau }+\frac{w_2}{\tau }+\frac{1}{\tau
   w_2}+\frac{1}{\tau  w_1}\right) \\
&\quad +x \left(\frac{2}{\tau
   ^2}+\frac{w_1^2}{\tau ^2}+\frac{w_2 w_1}{\tau ^2}+\frac{w_1}{\tau ^2
   w_2}+\frac{w_2^2}{\tau ^2}+\frac{1}{\tau ^2 w_2^2}+\frac{w_2}{\tau ^2
   w_1}+\frac{1}{\tau ^2 w_2 w_1}+\frac{1}{\tau ^2
   w_1^2}\right) \\
&\quad +O\left(x^{3/2}\right)
\end{aligned}
\end{align}
one can see that the $x^1$ term can be written as $x \tau^{-2} \chi^{Sp(2)}_\text{adj}$, which represents the BPS sector of the lowest component of the global current supermultiplet. Therefore, the enhanced IR symmetry is $Sp(2)$. We already denoted its Cartan as $U(1)_{B,1} \times U(1)_{B,2}$ with the fugacities $w_1,w_2$. For general $N_c = N_f = N$, one can see that the UV global symmetry $U(1)_T$ is enhanced to $Sp(N)$ in IR.\footnote{We adopt the 
convention that the rank of $Sp(N)$ is $N$.}

\subsubsection{$N_c < N_f < 2 N_c-1$}
Let us consider the $U(N_c)$ gauge theory with $N_f$ flavors where $N_c < N_f < 2 N_c-1$. Its dual theory is the $U(N_f-N_c)$ gauge theory with $N_f$ flavors and $2 N_c-N_f$ decoupled free twisted hypermultiplets. One can check that the dual theory is a ``good'' theory. Thus, we expect no additional IR symmetry emerges for the dual theory. Then we are able to write down the refined index of the dual theory by only using the UV content. It is given by
\begin{align}
I^{N_f-N_c,N_f}(x,t^{-1},\tau,w) \times \prod_{i = 1}^{2 N_c-N_f} I_\text{hyper}(x,\tau,w_i).
\end{align}
Again this refined index should be that of the original theory. Comparing with \eqref{eq:index duality}, the refined index is reduced to the partially refined index \eqref{eq:index duality} with $w_i \rightarrow w \tau^{2 N_c-N_f-2 i+1} x^{-(2 N_c-N_f-2 i+1)/2}$. Therefore, the UV charges and IR charges are identified as follows:
\begin{align}
R^{UV} &= R^{IR}-\frac{1}{2} \sum_{i = 1}^{2 N_c-N_f} (2 N_c-N_f-2 i+1) B_i^{IR}, \\
A^{UV} &= A^{IR}+\sum_{i = 1}^{2 N_c-N_f} (2 N_c-N_f-2 i+1) B_i^{IR}, \\
T^{UV} &= T^{IR}+\sum_{i = 1}^{2 N_c-N_f} B_i^{IR}.
\end{align}
This result is consistent with the general pattern observed in \cite{Bashkirov:2013dda}. The monopole operators $V_{i,\pm}$ in the $U(N_c)$ theory are mapped to either free twisted hypermultiplets or the monopole operators $\tilde V_{i,\pm}$ in the dual $U(N_f-N_c)$ theory:
\begin{align}
\begin{aligned}
(V_{i,+},V_{2 N_c-N_f-1-i,-}) &\leftrightarrow \text{free twisted hypers}, \\
V_{i,\pm} &\leftrightarrow \tilde V_{N_f-2 N_c+i,\pm}.
\end{aligned} \qquad
\begin{aligned}
i &= 0,\ldots,2 N_f-N_c-1 \\
i &= 2 N_c-N_f,\ldots,N_c
\end{aligned}
\end{align}
Recall that $T^{IR}$ was absent when $N_c = N_f$. For $N_c \neq N_f$, the UV global symmetry $SU(N_f) \times U(1)_T$ is enhanced to $SU(N_f) \times U(1)_T \times Sp(2 N_c-N_f)$ in IR. For instance, if we consider the $U(3)$ theory with four flavors, its refined index is obtained from the dual theory index as follows:
\begin{align}
\begin{aligned}
&\quad I^{1,3}(x,t^{-1},\tau,w) \times \prod_{i = 1}^{2} I_\text{hyper}(x,\tau,w_i) \\
&= 1+\sqrt{x} \left(\frac{w_1}{\tau }+\frac{w_2}{\tau }+\frac{1}{\tau
   w_2}+\frac{1}{\tau  w_1}\right) \\
&\quad +x \left(3 \tau ^2+\frac{t_2 \tau ^2}{t_1}+\frac{t_3 \tau ^2}{t_1}+\frac{t_3 \tau
   ^2}{t_2}+\frac{t_4 \tau ^2}{t_1}+\frac{t_4 \tau ^2}{t_2}+\frac{t_4 \tau
   ^2}{t_3}+\frac{t_1 \tau ^2}{t_2}+\frac{t_1 \tau ^2}{t_3}+\frac{t_2 \tau
   ^2}{t_3}+\frac{t_1 \tau ^2}{t_4}+\frac{t_2 \tau ^2}{t_4}+\frac{t_3 \tau
   ^2}{t_4}\right. \\
&\qquad \left.+\frac{3}{\tau
   ^2}+\frac{w_1^2}{\tau ^2}+\frac{w_2^2}{\tau ^2}+\frac{w_1 w_2}{\tau
   ^2}+\frac{w_2}{\tau ^2 w_1}+\frac{w_1}{\tau ^2 w_2}+\frac{1}{\tau ^2 w_1
   w_2}+\frac{1}{\tau ^2 w_1^2}+\frac{1}{\tau ^2
   w_2^2}\right)+O\left(x^{3/2}\right).
\end{aligned}
\end{align}
The $x^1$ term is written as $x (\tau^2 \chi^{SU(4)}_\text{adj}+\tau^{-2}+\tau^{-2} \chi^{Sp(2)}_\text{adj})$, which indicates that the global symmetry is $SU(4) \times U(1)_T \times Sp(2)$.

\section{3d $\mathcal N = 2$ Seiberg-like duality with an adjoint} \label{sec:N=2}
The second application of the factorization is the duality of 3d $\mathcal N = 2$ $U(N_c)_\kappa$ gauge theories with $N_f$ fundamental $Q_a$, $N_a$ antifundamental $\tilde Q_a$, one adjoint matter $X$ and the superpotential $W = \mathrm{tr} X^{n+1}$. The Chern-Simons coupling $\kappa$ should satisfy the condition $\kappa+\frac{N_f+N_a}{2} \in \mathbb Z$ due to the quantization of the effective CS coupling. Also we restrict our interest, as in section \ref{sec:factorization}, to the cases with $|\kappa| \leq \frac{|N_f-N_a|}{2}$. The theory has the global symmetry $SU(N_f) \times SU(N_a) \times U(1)_A \times U(1)_T$ as well as the $R$-symmetry $U(1)_R$. $U(1)_X$ doesn't exist due to the superpotential. The $U(1)_R$ charge of the adjoint chiral multiplet is fixed to $\frac{2}{n+1}$.

If $N_f = N_a$, it has been proposed that the theory has a Seiberg-like dual,  $U(n N_f-N_c)$ gauge theory with $N_f$ pairs of fundamental $q_{\tilde b}$ and antifundemental $\tilde q^a$, one adjoint $Y$, and $n ({N_f}^2+2)$ singlet matters ${M_i}_a^{\tilde b}$ and $V_{i,\pm}$ where $i = 0,\ldots,n-1$ \cite{Kim:2013cma}. The theory has the superpotential
\begin{align}
W = \mathrm{tr} Y^{n+1}+\sum_{i = 0}^{n-1} M_i \tilde q Y^{n-1-i} q+\sum_{i = 0}^{n-1} (V_{i,+} v_{n-1-i,-}+V_{i,-} v_{n-1-i,+})
\end{align}
where $v_{i,\pm}$'s are the monopole operators of the dual theory with the minimal fluxes. Let us call it KP duality. The global symmetry and charges are summarized in table \ref{tab:KP charges}.\begin{table}[tbp]
\centering
\begin{tabular}{|c|ccccc|}
\hline
 & $U(1)_R$ & $SU(N_f)_1$ & $SU(N_f)_2$ & $U(1)_A$ & $U(1)_T$ \\
\hline
$Q$ & $\Delta_Q$ & $\overline{\mathbf N_f}$ & $\mathbf 1$ & 1 & 0 \\
$\tilde Q$ & $\Delta_Q$ & $\mathbf 1$ & $\mathbf N_f$ & 1 & 0 \\
$X$ & $\frac{2}{n+1}$ & $\mathbf 1$ & $\mathbf 1$ & 0 & 0 \\
\hline
$q$ & $\frac{2}{n+1}-\Delta_Q$ & $\mathbf 1$ & $\overline{\mathbf N_f}$ & -1 & 0 \\
$\tilde q$ & $\frac{2}{n+1}-\Delta_Q$ & $\mathbf N_f$ & $\mathbf 1$ & -1 & 0 \\
$Y$ & $\frac{2}{n+1}$ & $\mathbf 1$ & $\mathbf 1$ & 0 & 0 \\
$M_i$ & $2 \Delta_Q+\frac{2 i}{n+1}$ & $\overline{\mathbf N_f}$ & $\mathbf N_f$ & 2 & 0 \\
$V_{i,\pm}$ & $\substack{(1-\Delta_Q) N_f\\-\frac{2}{n+1} (N_c-1-i)}$ & $\mathbf 1$ & $\mathbf 1$ & $-N_f$ & $\pm1$ \\
\hline
$v_{i,\pm}$ & $\substack{-(1-\Delta_Q) N_f\\+\frac{2}{n+1} (N_c+1+i)}$ & $\mathbf 1$ & $\mathbf 1$ & $N_f$ & $\pm1$ \\
\hline
\end{tabular}
\caption{\label{tab:KP charges} The global symmetry charges for the KP duality.}
\end{table}

\subsection{Generalization of KP duality to chiral-like theories}
In this subsection, we investigate the generalization of the KP duality for chiral-like theories, which may include the CS coupling under the condition $|\kappa| \leq \frac{|N_f-N_a|}{2}$.

Recall that the theory with $N_f=N_a$ and $\kappa=0$ has two $SU(N_f)$ global symmetries, which we denote by $SU(N_f)_1$ and $SU(N_f)_2$. The former rotates $Q$ while the latter rotates $\tilde Q$. Considering the combination with the axial symmetry $U(1)_A$ and the diagonal $U(1)_G$ of the gauge symmetry, the symmetries can be written as $SU(N_f)_1 \times SU(N_f)_2 \times U(1)_A \times U(1)_G \sim SU(N_f)_1 \times U(N_f)_2 \times U(1)_G$.\footnote{Here we specify the global symmetry and a part of the gauge symmetry since in the dual theory we have to consider the mixed Chern-Simons terms bewteen the global symmetry and the gauge symmetry
 $U(1)_G$.} Then we will consider a real mass deformation for $U(N_f)_2$. In particular,
denoting the Cartan subgroup of $U(N_f)_2$ by $\prod_{\tilde a' = 1}^{N_f} U(1)_{\tilde a'}$, we turn on real masses of $U(1)_{\tilde a'}$ for $\tilde a' = N_a+1,\ldots,N_f$ so that $N_f-N_a$ of the antifundamental matters are integrated out. The real mass corresponds to turning on the scalar vev for the background vector multiplet of interest. In this procedure, each charged massive fermion of mass $m$ generates a CS term at level $\Delta \kappa = \frac{1}{2} \mathrm{sign}(m)$ for the gauge symmetry. In fact, it can also generate a mixed CS term at level
\begin{align}
\Delta \kappa_{i j} = \frac{1}{2} q_i q_j \mathrm{sign}(m)
\end{align}
for each pair of abelian factors of the symmetries labeled by $i,j$. $q_i,q_j$ are corresponding abelian charges of the fermion. We want to avoid extra mixed CS terms associated with the residual global symmetries after integrating out the fermions. Thus, we first redefine abelian global symmetries such that the massive fermions are not charged under them:
\begin{align}
R^\text{new} &= R-(\Delta_Q-1) \sum_{\tilde a' = N_a+1}^{N_f} F_{\tilde a'}, \label{eq:new symmetries} \\
A^\text{new} &= A-\sum_{\tilde a' = N_a+1}^{N_f} F_{\tilde a'}
\end{align}
where $R,A$ are the $U(1)_R,U(1)_A$ charges and each $F_{\tilde a'}$ is the $U(1)_{\tilde a'}$ charge. Then the new charges of $Q$ and $\tilde Q$ are given in table \ref{tab:new charges}.
\begin{table}[tbp]
\centering
\begin{tabular}{|c|cccccc|}
\hline
 & $U(1)_R^\text{new}$ & $SU(N_f)$ & $SU(N_a)$ & $U(1)_{\tilde a'}$ & $U(1)_A^\text{new}$ & $U(1)_T$ \\
\hline
$Q$ & $\Delta_Q$ & $\overline{\mathbf N_f}$ & $\mathbf 1$ & 0 & 1 & 0 \\
$\tilde Q^{\tilde b}$ & $\Delta_Q$ & $\mathbf 1$ & $\mathbf{N_a}$ & 0 & 1 & 0 \\
$\tilde Q^{\tilde b'}$ & 1 & $\mathbf 1$ & $\mathbf 1$ & $\delta_{\tilde a' \tilde b'}$ & 0 & 0 \\
$X$ & $\frac{2}{n+1}$ & $\mathbf 1$ & $\mathbf 1$ & 0 & 0 & 0 \\
\hline
$q_{\tilde b}$ & $\frac{2}{n+1}-\Delta_Q$ & $\mathbf 1$ & $\overline{\mathbf{N_a}}$ & 0 & -1 & 0 \\
$q_{\tilde b'}$ & $\frac{2}{n+1}-1$ & $\mathbf 1$ & $\mathbf 1$ & $-\delta_{\tilde a' \tilde b'}$ & 0 & 0 \\
$\tilde q$ & $\frac{2}{n+1}-\Delta_Q$ & $\mathbf N_f$ & $\mathbf 1$ & 0 & -1 & 0 \\
$Y$ & $\frac{2}{n+1}$ & $\mathbf 1$ & $\mathbf 1$ & 0 & 0 & 0 \\
$M_i{}_a^{\tilde b}$ & $2 \Delta_Q+\frac{2 i}{n+1}$ & $\overline{\mathbf N_f}$ & $\mathbf{N_a}$ & 0 & 2 & 0 \\
$M_i{}_a^{\tilde b'}$ & $\Delta_Q+1+\frac{2 i}{n+1}$ & $\overline{\mathbf N_f}$ & $\mathbf 1$ & $\delta_{\tilde a' \tilde b'}$ & 1 & 0 \\
$V_{i,\pm}$ & $\substack{\frac{1}{2} (1-\Delta_Q) (N_f+N_a)\\-\frac{2}{n+1} (N_c-1-i)}$ & $\mathbf 1$ & $\mathbf 1$ & $-\frac{1}{2}$ & $-\frac{N_f+N_a}{2}$ & $\pm1$ \\
\hline
$v_{i,\pm}$ & $\substack{-\frac{1}{2} (1-\Delta_Q) (N_f+N_a)\\+\frac{2}{n+1} (N_c+1+i)}$ & $\mathbf 1$ & $\mathbf 1$ & $\frac{1}{2}$ & $N_f$ & $\pm1$ \\
\hline
\end{tabular}
\caption{\label{tab:new charges} The new symmetry charges for the KP duality. The indices $\tilde a',\tilde b'$ and $\tilde b$ run over $\tilde a',\tilde b' = N_a+1,\ldots,N_f$ and $\tilde b = 1,\ldots,N_a$. We will turn on real masses for $\prod_{\tilde a' = N_a+1}^{N_f} U(1)_{\tilde a'}$.}
\end{table}
In this way, one can avoid the mixed CS terms associated with the global symmetries $U(1)_R^\text{new} \times U(1)_A^\text{new} \times U(1)_T \times SU(N_f)_1 \times SU(N_a)$. However, the mixed CS terms associated with the abelian factors of $SU(N_f)_2$ are unavoidable. The massive fermions generate extra mixed CS terms between $U(1)_G$ and $U(1)_{\tilde a'}$'s, which look like shifts of the FI coupling in the low-energy theory.\footnote{The N=2 SUSY completion of
the mixed Chern-Simons term $A_{\tilde{a}'} \wedge F_{U(1)_G}$ has the term $\sigma_{\tilde{a}'} D_{U(1)_G} \sim m D_{U(1)_G} $ where $A_{\tilde{a}'}$ and $\sigma_{\tilde{a}'}$ are respectively 
the vector potential and scalar of $N=2$ background vector multiplet of $U(1)_{\tilde a'}$.} Thus, we introduce a bare FI coupling to the theory so that its low-energy theory doesn't have the FI term. In particular, if we turn on $N_+$ positive real masses and $N_-$ negative real masses for antifundamental matters, the low-energy effective theory is a $U(N_c)_{\kappa,\zeta}$ theory with $N_a = N_f-N_+-N_-$ antifundamental matters as well as the $N_f$ fundamental and one adjoint matters. The CS and FI couplings $\kappa,\zeta$ are given by
\begin{align}
\kappa = \frac{1}{2} (N_+-N_-), \qquad \zeta = \zeta_0-\frac{1}{2} \left(\sum_{\tilde b' = N_a+1}^{N_a+N_+} m_{\tilde b'}-\sum_{\tilde b' = N_a+N_++1}^{N_f} m_{\tilde b'}\right)
\end{align}
where $\zeta_0$ is the bare FI coupling and $m_{\tilde b'}$ is real mass for $U(1)_{\tilde b'}$, which is positive for $\tilde b' = N_a+1,\ldots,N_a+N_+$ or negative for $\tilde b' = N_a+N_++1,\ldots,N_f$. Thus, by taking a bare FI coupling
\begin{align}
\zeta_0 = \frac{1}{2} \left(\sum_{\tilde b' = N_a+1}^{N_a+N_+} m_{\tilde b'}-\sum_{\tilde b' = N_a+N_++1}^{N_f} m_{\tilde b'}\right),
\end{align}
the low-energy effective theory has the vanishing FI term. There are also mixed CS terms among $U(1)_{\tilde a'}$'s, whose effect is trivial when the background flux for the symmetries is absent. Note that in this low-energy theory, a gauge invariant bare monopole operator exists only when the effective CS level at $\sigma \rightarrow \pm\infty$,
\begin{align}
\kappa_\text{eff}(\sigma \rightarrow \pm\infty) = \kappa\pm\frac{N_f-N_a}{2} = \pm N_\pm,
\end{align}
vanishes; i.e., $N_\pm = 0$. Alternatively the monopole operators for chiral-like theories carry a nonzero zero-point charge, which effectively
changes the Chern-Simons level \cite{Imamura:2011su}.

The real masses in the original theory are translated to real masses in the dual theory dictated by the global symmetries. Table \ref{tab:new charges} again shows charges under the new symmetries \eqref{eq:new symmetries} in the dual theory. One can see that the fundamental matter $q_{\tilde b'}$ has real mass $-m_{\tilde b'}$, which is negative for $\tilde b' = N_a+1,\ldots,N_a+N_+$ or positive for $\tilde b' = N_a+N_++1,\ldots,N_f$. Similarly, the singlet matter $M_i{}_a^{\tilde b'}$ has real mass $m_{\tilde b'}$. On the other hand, real masses of $V_\pm$ also include contributions of the FI coupling:
\begin{align}
m_+ &= -\frac{1}{2} \sum_{\tilde b' = N_a+1}^{N_f} m_{\tilde b'}+\zeta_0 = -\sum_{\tilde b' = N_a+N_++1}^{N_f} m_{\tilde b'},  \label{mass1}\\
m_- &= -\frac{1}{2} \sum_{\tilde b' = N_a+1}^{N_f} m_{\tilde b'}-\zeta_0 = -\sum_{\tilde b' = N_a+1}^{N_a+N_+} m_{\tilde b'},
\end{align}
each of which vanishes for $N_- = 0$ or for $N_+ = 0$ respectively.\footnote{ $-\frac{1}{2} \sum_{\tilde b' = N_a+1}^{N_f} m_{\tilde b'}$ of eq. (\ref{mass1}) corresponds to the scalar vev of $U(1)_{A^{new}}$ and $\zeta_0$ is the scalar vev of $U(1)_T$ vector multiplet. We have the BF term
$A_{U(1)_T}\wedge F_{U(1)_G}$ so that monopole operator is charged under $U(1)_T$.} This is consistent with the fact that on the original $U(N_c)$ theory side, a gauge invariant bare monopole operator exists only when $N_\pm = 0$.
The CS and FI couplings in the low-energy theory are
\begin{gather}
\kappa^\text{dual} = -\frac{1}{2} (N_+-N_-) = -\kappa, \\
\zeta^\text{dual} = -\zeta_0+\frac{1}{2} \left(\sum_{\tilde b' = N_a+1}^{N_a+N_+} m_{\tilde b'}-\sum_{\tilde b' = N_a+N_++1}^{N_f} m_{\tilde b'}\right) = 0.
\end{gather}
Thus, integrating out the massive fields, the low-energy effective theory of the dual theory is a $U(n N_f-N_c)_{-\kappa}$ theory with $N_a$ fundamental $q$, $N_f$ antifundamental $\tilde q$, one adjoint $Y$, $n N_f N_a$ singlets $M_i$ and possibly $n$ singlet matters $V_{i,+}$ or $V_{i,-}$ depending on $\kappa$. Again there exists a bare monopole operator in the dual low-energy theory if $N_+ = 0$ or $N_- = 0$. In particular, when $N_+ = 0$, both $V_{i,+}$ and $v_{i,-}$ exist, so the low-energy theory inherits the superpotential terms
\begin{align} \label{eq:superpotential+}
\sum_{i = 0}^{n-1} V_{i,+} v_{n-1-i,-}.
\end{align}
Those superpotential terms are crucial for the duality because they lift the monopole operators $v_{i,-}$, which do not appear in the original theory.
Likewise, when $N_- = 0$, both $V_{i,-}$ and $v_{i,+}$ exit, so the low-energy theory inherits the superpotential terms
\begin{align} \label{eq:superpotential-}
\sum_{i = 0}^{n-1} V_{i,-} v_{n-1-i,+},
\end{align}
which lift the monopole operators $v_{i,+}$.

In contrast to the original $U(N_c)$ theory, one should note that there are massive fermions charged under $U(1)_R^\text{new}$ in the dual theory. They generate a mixed CS term between $U(1)_R$ and $U(1)_G$ at level
\begin{align} \label{eq:CS RG}
\kappa_{RG} = \frac{n}{n+1} (N_+-N_-) = \frac{2 \kappa n}{n+1}
\end{align}
since the R charge of the fermion partner of $q_{\tilde{a}'}$ is given by $2\frac{n}{n+1}-2$. The additional minus sign is due to the negative real mass
for $q_{\tilde{a}'}$.

This mixed CS term shifts the $R$-charges of monopole operators by
\begin{align} \label{eq:shift}
\Delta R = \frac{2 \kappa n}{n+1} T
\end{align}
where $T$ is the $U(1)_T$ charge.\footnote{Using the operator-state correspondence of the conformal field theory, this can be understood  from the Gauss constraint  for $U(1)_G$ in the presence of
mixed Chern-Simons term $A_{R} \wedge F_{U(1)_G}$ on $R \times S^2$. Gauss constraint has the form $k$ Flux$_{U(1)_G}=$  R charge
. Since $A_{R}$ is not dynamical, we do not have to impose the Gauss constraint. The constraint simply dictates the R charge contribution carried by the monopole operator. } This shift is crucial for the duality. Let us consider the $N_+ = 0$ case first. In that case, the dual low-energy theory has the gauge invariant monopole operators $v_{i,-} \sim X^i \left|1,\ldots,0\right>$. Without the shift \eqref{eq:shift}, the $R$-charge of a bare monopole state of flux $m$ is determined by \cite{Kim:2009wb,Imamura:2011su}
\begin{align} \label{eq:bare monopole R-charge}
\Delta(m) &= \frac{1}{2} \sum_{\Phi} (1-\Delta_\Phi) \sum_{\rho \in R_\Phi} |\rho(m)|-\frac{1}{2} \sum_{\alpha \in G} |\alpha(m)|
\end{align}
where $\Phi$ denotes every charged chiral multiplet, which is in the representation $R_\Phi$. $\rho$ is a weight of the representation $R_\Phi$ and $\alpha$ is a root of the gauge group $G$. \eqref{eq:bare monopole R-charge} implies that the $R$-charges of $v_{i,-}$ are
\begin{align} \label{eq:R0}
R_{\kappa_{RG} = 0}= -\frac{1}{2} \left(1-\Delta_Q\right) (N_f+N_a)+\frac{2}{n+1} (N_c+1+i)-\frac{n}{n+1} (N_f-N_a).
\end{align}
With those $R$-charges of $v_{i,0}$, the superpotential terms \eqref{eq:superpotential+} are not available because their $R$-charge, $2-\frac{n}{n+1} (N_f-N_a) \neq 2$, is anomalous. Indeed, the shift \eqref{eq:shift}
\begin{align}
\Delta R = \frac{n}{n+1} N_-
\end{align}
exactly cancels the last term in \eqref{eq:R0} and compensates the anomalous $R$-charge of the superpotential terms. Thus, the superpotential terms \eqref{eq:superpotential+} are available only in the presence of the mixed CS term \eqref{eq:CS RG}. For the same reason, when $N_- = 0$, the superpotential terms \eqref{eq:superpotential-} are only available in the presence of the mixed CS term \eqref{eq:CS RG}, which shifts the $R$-charges of $v_{i,+}$. The shifted $R$-charges are in table \ref{tab:new charges}.

In conclusion, assuming $N_f > N_a$ we propose that
\begin{itemize}
\item $U(N_c)_\kappa$ theory with $N_f$ fundamental $Q_a$, $N_a$ antifundamental $\tilde Q^{\tilde b}$, one adjoint $X$ and the superpotential $W = X^{n+1}$
\end{itemize}
has a Seiberg-like dual description
\begin{itemize}
\item $U(n N_f-N_c)_{-\kappa}$ theory with $N_a$ fundamental $q_{\tilde b}$, $N_f$ antifundamental $\tilde q^a$, one adjoint $Y$, and
\begin{itemize}
\item $n N_f N_a$ singlet matters ${M_i}_a^{\tilde b}$ with $i = 0,\ldots,n-1$ and the superpotential
\begin{align} \label{eq:superpotential 1}
W = \mathrm{tr} Y^{n+1}+\sum_{i = 0}^{n-1} M_i \tilde q Y^{n-1-i} q
\end{align}
if $|\kappa| < \frac{N_f-N_a}{2}$.
\item $n N_f N_a+n$ singlet matters ${M_i}_a^{\tilde b}$ and $V_{i,+}$ with $i = 0,\ldots,n-1$ and the superpotential
\begin{align} \label{eq:superpotential 2}
W = \mathrm{tr} Y^{n+1}+\sum_{i = 0}^{n-1} M_i \tilde q Y^{n-1-i} q+\sum_{i = 0}^{n-1} V_{i,+} v_{n-1-i,-}
\end{align}
if $\kappa = -\frac{N_f-N_a}{2}$.
\item $n N_f N_a+n$ singlet matters ${M_i}_a^{\tilde b}$ and $V_{i,-}$ with $i = 0,\ldots,n-1$ and the superpotential
\begin{align} \label{eq:superpotential 3}
W = \mathrm{tr} Y^{n+1}+\sum_{i = 0}^{n-1} M_i \tilde q Y^{n-1-i} q+\sum_{i = 0}^{n-1} V_{i,-} v_{n-1-i,+}
\end{align}
if $\kappa = \frac{N_f-N_a}{2}$.
\end{itemize}
\end{itemize}
The global symmetry and charges are summarized in table \ref{tab:global charges 2}.
\begin{table}[tbp]
\centering
\begin{tabular}{|c|ccccc|}
\hline
 & $U(1)_R$ & $SU(N_f)$ & $SU(N_a)$ & $U(1)_A$ & $U(1)_T$ \\
\hline
$Q$ & $\Delta_Q$ & $\overline{\mathbf N_f}$ & $\mathbf 1$ & 1 & 0 \\
$\tilde Q$ & $\Delta_Q$ & $\mathbf 1$ & $\mathbf N_a$ & 1 & 0 \\
$X$ & $\frac{2}{n+1}$ & $\mathbf 1$ & $\mathbf 1$ & 0 & 0 \\
\hline
$q$ & $\frac{2}{n+1}-\Delta_Q$ & $\mathbf 1$ & $\overline{\mathbf N_a}$ & -1 & 0 \\
$\tilde q$ & $\frac{2}{n+1}-\Delta_Q$ & $\mathbf N_f$ & $\mathbf 1$ & -1 & 0 \\
$Y$ & $\frac{2}{n+1}$ & $\mathbf 1$ & $\mathbf 1$ & 0 & 0 \\
$M_i$ & $2 \Delta_Q+\frac{2 i}{n+1}$ & $\overline{\mathbf N_f}$ & $\mathbf N_a$ & 2 & 0 \\
$V_{i,\pm}$ & $\substack{\frac{1}{2} (1-\Delta_Q) (N_f+N_a)\\-\frac{2}{n+1} (N_c-1-i)}$ & $\mathbf 1$ & $\mathbf 1$ & $-\frac{N_f+N_a}{2}$ & $\pm1$ \\
\hline
$v_{i,\pm}$ & $\substack{-\frac{1}{2} (1-\Delta_Q) (N_f+N_a)\\+\frac{2}{n+1} (N_c+1+i)}$ & $\mathbf 1$ & $\mathbf 1$ & $\frac{N_f+N_a}{2}$ & $\pm1$ \\
\hline
\end{tabular}
\caption{\label{tab:global charges 2} The global symmetry charges for the proposed duality for chiral-like theories. The monopole operators $V_{i,\pm}$ and $v_{i,\mp}$ only appear when $\kappa\pm\frac{N_f-N_a}{2} = 0$.}
\end{table}
The dual theory also has the mixed CS term at level \eqref{eq:CS RG} between the $U(1)_R$ $R$-symmetry and the diagonal $U(1)_G$ of the gauge symmetry.\footnote{There are also mixed CS terms among the global symmetries not associated with the gauge symmetry. However, their effect is trivial when the background flux for the symmetries is absent.} The superpotentials (\ref{eq:superpotential 1}-\ref{eq:superpotential 3}) are crucial for the duality because they lift dual theory operators $\tilde q Y^i q$ and $v_{i,\pm}$, which do not appear in the original theory. Note that this duality is also a generalization of the Seiberg-like duality for chiral-like theories without an adjoint matter \cite{Benini:2011mf}. If $n = 1$, the duality we propose is reduced to that of \cite{Benini:2011mf} by integrating out the adjoint matter.

The superconformal indices of KP duality pairs were computed as power series around $x = 0$ in \cite{Kim:2013cma}. It was checked for several values of $N_c,N_f,n$ that those two indices coincide, which provides strong evidence of the duality. However, such comparisons were restricted to  the cases satisfying the condition
\begin{align} \label{eq:positive R-charge}
N_f-\frac{2}{n+1} (N_c-1) > 0
\end{align}
because only in those cases, the superconformal indices are analytic at $x = 0$ such that the power series of the indices around $x = 0$ exist. One might think that such cases are enough because only in those case, the monopole operators of the theory have the positive UV $R$-charges. However, as we have seen in the previous $\mathcal N = 4$ example, the UV $R$-symmetry can be corrected by accidental IR symmetries such that the IR $R$-charges are larger than or equal to 1/2, which do not violate the unitarity. Therefore, we need a tool for testing the cases not satisfying the condition \eqref{eq:positive R-charge}. Indeed, the factorized index we obtained is such a tool since it doesn't require analyticity at $x = 0$. We will see that one can compare the factorized indices of a duality pair even if the condition \eqref{eq:positive R-charge} is not satisfied.

Furthermore, we investigate exact relations of the factorized indices for the duality we propose above for chiral-like theories. Those relations are nontrivial evidence of the proposed duality.

\subsection{SCI under duality}
In the presence of the superpotential $W = \mathrm{tr} X^{n+1}$, the adjoint chiral multiplet $X$ has the $R$-charge $\frac{2}{n+1}$ and no other global charge. Let us call that value of the $R$-charge $\delta$. Therefore, in order to obtain the factorized index for this case, we need to substitute $\upsilon = x^\frac{2}{n+1} = x^\delta$. Then one can easily see that $I_\text{pert}$ vanishes if $p_a > n$ in eq. (\ref{facteq}). Thus, the superconformal index for this case is given by
\begin{align}
I(x,t,\tilde t,\tau,x^\delta,w) = \sum_{\substack{0 \leq p_a \leq n,\\\sum_a p_a = N_c}} I_\text{pert}^{(p_a)}(x,t,\tilde t,\tau,x^\delta) Z_\text{vortext}^{(p_a)}(x,t,\tilde t,\tau,x^\delta,\mathfrak w) Z_\text{antivortex}^{(p_a)}(x,t,\tilde t,\tau,x^\delta, \mathfrak w)\end{align}
where $\mathfrak w = (-1)^{-\kappa-\frac{N_f-N_a}{2}} w$. $p_a$'s are constrained such that $0 \leq p_a \leq n$ and $\sum_{a = 1}^{N_f} p_a = N_c$. Each component is also given by
\begin{align}
\begin{aligned}
&\quad I_\text{pert}^{(p_a)}(x,t = e^{i M},\tilde t,\tau,x^\delta = e^{-\delta \gamma}) \\
&= \left(\prod_{a,b = 1}^{N_f} \prod_{q = 1}^{p_a} \prod_{\substack{r = 1 \\ (\neq q \text{ if } a = b)}}^{p_b} 2 \sinh \frac{i M_{a}-i M_{b}-\delta \gamma (q-r)}{2}\right) \left(\prod_{a,b = 1}^{N_f} \prod_{q = 1}^{p_a} \prod_{r = 1}^{p_b} \frac{\left(t_{a} t_{b}^{-1} x^{\delta (q-r-1)+2};x^2\right)_\infty}{\left(t_{a}^{-1} t_{b} x^{\delta (-q+r+1)};x^2\right)'_\infty}\right) \\
&\quad \times \left(\prod_{a = 1}^{N_f} \prod_{q = 1}^{p_a} \frac{\prod_{b = 1}^{N_f} \left(t_a t_b^{-1} x^{\delta (q-1)+2};x^2\right)_\infty}{\prod_{b = 1}^{N_a} \left(t_a \tilde t_b \tau^2 x^{\delta (q-1)};x^2\right)_\infty} \frac{\prod_{b = 1}^{N_a} \left(t_a^{-1} \tilde t_b^{-1} \tau^{-2} x^{\delta (-q+1)+2};x^2\right)_\infty}{\prod_{b = 1}^{N_f} \left(t_a^{-1} t_b x^{\delta (-q+1)};x^2\right)'_\infty}\right),
\end{aligned}
\end{align}
\begin{align}
Z_\text{vortex}^{(p_a)}(x,t,\tilde t,\tau,x^\delta,\mathfrak w) &= \sum_{\mathfrak n_j \geq 0} \mathfrak w^{\sum_{j = 1}^{N_c} \sum_{n = 0}^{l_j-1} \mathfrak n_j^n} \mathfrak Z^{(p_a)}_{(\mathfrak n_j)}(x,t,\tilde t,\tau,x^\delta), \\
Z_\text{antivortex}^{(p_a)}(x,t,\tilde t,\tau,x^\delta,\mathfrak w) &= \sum_{\bar{\mathfrak n}_j \geq 0} \mathfrak w^{-\sum_{j = 1}^{N_c} \sum_{n = 0}^{l_j-1} \bar{\mathfrak n}_j^n} \mathfrak Z^{(p_a)}_{(\bar{\mathfrak n}_j)}(x^{-1},t^{-1},\tilde t^{-1},\tau^{-1},x^{-\delta}),
\end{align}
\begin{align}
\begin{aligned}
&\quad \mathfrak Z^{(p_a)}_{(\mathfrak n_j)}(x = e^{-\gamma},t = e^{i M},\tilde t = e^{i \tilde M},\tau = e^{i \mu},x^\delta = e^{-\delta \gamma}) \\
&= e^{-S^{(p_a)}_{(\mathfrak n_j)}(x,t,\tau,\upsilon)} \left(\prod_{a,b = 1}^{N_f} \prod_{q = 1}^{p_a} \prod_{\substack{r = 1 \\ (\neq q \text{ if } a = b)}}^{p_b} \prod_{k = 1}^{\sum_{n = 1}^r \mathfrak n_{(b,n)}} \frac{\sinh \frac{i M_a-i M_b-\delta \gamma (q-r)+2 \gamma k}{2}}{\sinh \frac{i M_a-i M_b-\delta \gamma (q-r)+ 2 \gamma (k-1-\sum_{n = 1}^q \mathfrak n_{(a,n)})}{2}}\right) \\
&\quad \times \left(\prod_{a,b = 1}^{N_f} \prod_{q = 1}^{p_a} \prod_{\substack{r = 1 \\ (\neq q \text{ if } a = b)}}^{p_b} \prod_{k = 1}^{\sum_{n = 1}^r \mathfrak n_{(b,n)}} \frac{\sinh \frac{i M_a-i M_b-\delta \gamma (q-r-1)+2 \gamma (k-1-\sum_{n = 1}^q \mathfrak n_{(a,n)})}{2}}{\sinh \frac{i M_a-i M_b-\delta \gamma (q-r+1)+2 \gamma k}{2}}\right) \\
&\quad \times \left(\prod_{b = 1}^{N_f} \prod_{r = 1}^{p_b} \prod_{k = 1}^{\sum_{n = 1}^r \mathfrak n_{(b,n)}} \frac{\prod_{a = 1}^{N_a} \sinh \frac{-i \tilde M_a-i M_b-2 i \mu+\delta \gamma (r-1)+2 \gamma (k-1)}{2}}{\prod_{a = 1}^{N_f} \sinh \frac{i M_a-i M_b+\delta \gamma (r-1)+2 \gamma k}{2}}\right),
\end{aligned}
\end{align}
where
\begin{align}
e^{-S^{(p_a)}_{(\mathfrak n_j)}(x,t,\tau,\upsilon)} = \prod_{b = 1}^{N_f} \prod_{r = 1}^{p_b} \left(t_b \tau \upsilon^{r-1} x^{\sum_{n = 1}^r \mathfrak n_{(b,n)}}\right)^{\kappa \sum_{n = 1}^r \mathfrak n_{(b,n)}}.
\end{align}
$\mathfrak n_{(a,q)}$ is a shorthand notation for $\mathfrak n_{q+\sum_{b = 1}^{a-1} p_b}$. Again the prime symbol indicates that the zero factor in the $q$-Pochhammer symbol is omitted.

Now we attempt to rephrase the index agreement for a KP duality pair in terms of the factorized index. Each component is mapped under duality as follows:
\begin{align}
&\begin{aligned} \label{eq:N=2 pert duality 0}
&\quad I_\text{pert}^{(p_a),N_c,N_f,N_f}(x,t,\tilde t,\tau) \\
&= I_\text{pert}^{(n-p_a),N_f-N_c,N_f,N_f}(x,t^{-1},\tilde t^{-1},\tau^{-1} x^\delta) \times \prod_{a,b = 1}^{N_f} \prod_{q = 1}^n \frac{Z_\text{chiral}(t_a \tilde t_b \tau^2 x^{\delta (q-1)})}{Z_\text{chiral}(t_a^{-1} \tilde t_b^{-1} \tau^{-2} x^{2-\delta (q-1)})},
\end{aligned} \\
&\begin{aligned} \label{eq:N=2 vpf duality 0}
&\quad Z_\text{vortex}^{(p_a),N_c,N_f,N_f}(x,t,\tilde t,\tau,w) \\
&= Z_\text{antivortex}^{(n-p_a),N_f-N_c,N_f,N_f}(x,t^{-1},\tilde t^{-1},\tau^{-1} x^\delta,w^{-1}) \times \prod_{i = 1}^{n} \frac{Z_\text{chiral}(x,w \tau^{-N_f} x^{\Delta_i})}{Z_\text{chiral}(x,w \tau^{N_f} x^{2-\Delta_i})},
\end{aligned} \\
&\begin{aligned} \label{eq:N=2 avpf duality 0}
&\quad Z_\text{antivortex}^{(p_a),N_c,N_f,N_f}(x,t,\tilde t,\tau,w) \\
&= Z_\text{vortex}^{(n-p_a),N_f-N_c,N_f,N_f}(x,t^{-1},\tilde t^{-1},\tau^{-1} x^\delta,w^{-1}) \times \prod_{i = 1}^{n} \frac{Z_\text{chiral}(x,w^{-1} \tau^{-N_f} x^{\Delta_i})}{Z_\text{chiral}(x,w^{-1} \tau^{N_f} x^{2-\Delta_i})}
\end{aligned}
\end{align}
where $\Delta_i = N_f-\frac{2}{n+1} (N_c-1-i)$, which is the $R$-charge of a monopole operator of the original theory. $Z_\text{chiral}$ is defined by
\begin{align}
Z_\text{chiral}(x,w) = \frac{1}{\left(w;x^2\right)_\infty} = \mathrm{PE}\left[\frac{w}{1-x^2}\right]
\end{align}
such that the index of a singlet chiral multiplet is written in terms of $Z_\text{chiral}$ as follows:
\begin{align}
I_\text{chiral}(x,w) = Z_\text{chiral}(x,w) \times Z_\text{chiral}(x^{-1},w^{-1}).
\end{align}
The generalization of \eqref{eq:N=2 pert duality 0} for the chiral version of the KP duality is straightforward. We will provide its explicit form and proof shortly. For \eqref{eq:N=2 vpf duality 0} and \eqref{eq:N=2 avpf duality 0}, we will show that the generalizations of them can be obtained by examining large mass behavior of the vortex partition function. Those identities together imply the index agreement for the duality we propose for chiral-like theories.

From the duality we propose, we expect \eqref{eq:N=2 pert duality 0} is generalized as follows:
\begin{align}
\begin{aligned} \label{eq:N=2 pert duality}
&\quad I_\text{pert}^{(p_a),N_c,N_f,N_a}(x,t,\tilde t,\tau) \\
&= I_\text{pert}^{(n-p_a),N_f-N_c,N_f,N_a}(x,t^{-1},\tilde t^{-1},\tau^{-1} x^\delta) \times \prod_{a = 1}^{N_f} \prod_{b = 1}^{N_a} \prod_{q = 1}^n \frac{Z_\text{chiral}(t_a \tilde t_b \tau^2 x^{\delta (q-1)})}{Z_\text{chiral}(t_a^{-1} \tilde t_b^{-1} \tau^{-2} x^{2-\delta (q-1)})}.
\end{aligned}
\end{align}
In order to show the generalized identity \eqref{eq:N=2 pert duality}, we start from noticing that the following identity holds:
\begin{align}
\begin{aligned}
\left(\prod_{a,b = 1}^{N_f} \prod_{q = 1}^{p_a} \prod_{r = 0}^{p_b-1} \frac{\left(t_{a} t_{b}^{-1} x^{\delta (q-r-1)+2};x^2\right)_\infty}{\left(t_{a}^{-1} t_{b} x^{\delta (-q+r+1)};x^2\right)'_\infty}\right) &= \left(\prod_{a,b = 1}^{N_f} \prod_{q = 1}^{p_a} \prod_{r = 1}^{p_b} \frac{\left(t_{a} t_{b}^{-1} x^{\delta (q-r)+2};x^2\right)_\infty}{\left(t_{a}^{-1} t_{b} x^{\delta (-q+r)};x^2\right)'_\infty}\right) \\
&= \left(\prod_{a,b = 1}^{N_f} \prod_{q = 1}^{p_a} \prod_{\substack{r = 1 \\ (\neq q \text{ if } a = b)}}^{p_b} 2 \sinh \frac{i M_{a}-i M_{b}-\delta \gamma (q-r)}{2}\right)^{-1}.
\end{aligned}
\end{align}
It cancels out the first factor of $I_\text{pert}^{(p_a)}$ so that the remaining factors are simply given by
\begin{align}
\left(\prod_{a,b = 1}^{N_f} \prod_{q = 1}^{p_a} \frac{\left(t_{a} t_{b}^{-1} x^{\delta (q-p_b-1)+2};x^2\right)_\infty}{\left(t_{a}^{-1} t_{b} x^{\delta (-q+p_b+1)};x^2\right)_\infty}\right) \left(\prod_{a = 1}^{N_f} \prod_{b = 1}^{N_a} \prod_{q = 1}^{p_a} \frac{\left(t_a^{-1} \tilde t_b^{-1} \tau^{-2} x^{\delta (-q+1)+2)};x^2\right)_\infty}{\left(t_a \tilde t_b \tau^2 x^{\delta (q-1)};x^2\right)_\infty}\right).
\end{align}
Then let us examine each factor. The first factor can be written as
\begin{align}
\begin{aligned}
&\quad \left(\prod_{a,b = 1}^{N_f} \prod_{q = 1}^{p_a} \frac{\left(t_{a} t_{b}^{-1} x^{\delta (q-p_b-1)+2};x^2\right)_\infty}{\left(t_{a}^{-1} t_{b} x^{\delta (-q+p_b+1)};x^2\right)_\infty}\right) \\
&= \left(\prod_{a,b = 1}^{N_f} \prod_{q = p_a-n+1}^{p_a} \frac{\left(t_{a} t_{b}^{-1} x^{\delta (q-p_b-1)+2};x^2\right)_\infty}{\left(t_{a}^{-1} t_{b} x^{\delta (-q+p_b+1)};x^2\right)_\infty}\right) \left(\prod_{a,b = 1}^{N_f} \prod_{q = p_a-n+1}^{0} \frac{\left(t_{a}^{-1} t_{b} x^{\delta (-q+p_b+1)};x^2\right)_\infty}{\left(t_{a} t_{b}^{-1} x^{\delta (q-p_b-1)+2};x^2\right)_\infty}\right) \\
&= \left(\prod_{a,b = 1}^{N_f} \prod_{q = 1}^{n-p_a} \frac{\left(t_{a}^{-1} t_{b} x^{\delta (q-n+p_b-1)+2};x^2\right)_\infty}{\left(t_{a} t_{b}^{-1} x^{\delta (-q+n-p_b+1)};x^2\right)_\infty}\right)
\end{aligned}
\end{align}
where it is used that the first factor after the first equality is simply 1. It shows that the factor is invariant under the change $t_a \rightarrow t_a^{-1}$ and $p_a \rightarrow n-p_a$. This is a crucial feature when we match the indices of a duality pair. Next the second factor is written as
\begin{align}
\begin{aligned}
&\quad \left(\prod_{a = 1}^{N_f} \prod_{b = 1}^{N_a} \prod_{q = 1}^{p_a} \frac{\left(t_a^{-1} \tilde t_b^{-1} \tau^{-2} x^{\delta (-q+1)+2)};x^2\right)_\infty}{\left(t_a \tilde t_b \tau^2 x^{\delta (q-1)};x^2\right)_\infty}\right) \\
&= \left(\prod_{a = 1}^{N_f} \prod_{b = 1}^{N_a}  \prod_{q = 1}^n \frac{\left(t_a^{-1} \tilde t_b^{-1} \tau^{-2} x^{\delta (-q+1)+2)};x^2\right)_\infty}{\left(t_a \tilde t_b \tau^2 x^{\delta (q-1)};x^2\right)_\infty}\right) \left(\prod_{a = 1}^{N_f} \prod_{b = 1}^{N_a}  \prod_{q = 1}^{n-p_a} \frac{\left(t_a \tilde t_b \tau^2 x^{\delta (-q-1)+2};x^2\right)_\infty}{\left(t_a^{-1} \tilde t_b^{-1} \tau^{-2} x^{\delta (q+1)};x^2\right)_\infty}\right)
\end{aligned}
\end{align}
where the first factor of the right hand side is nothing but the contribution of $n N_f N_a$ singlets ${M_i}_a^{\tilde b}$. Combining the results, we prove the identity \eqref{eq:N=2 pert duality}, which supports our proposal.

Now, let us examine large mass behavior of the vortex partition function. Especially we are interested in the cases that real mass of an antifundamental matter goes to $\pm\infty$. Thus, let us choose the $N_f$-th antifundamental matter and take its mass large. We first observe that $\mathfrak Z_{(n_j)}^{(p_a)}$ has asymptotic behavior
\begin{align}
\begin{aligned}
&\mathfrak Z^{(p_a),N_c,N_f,N_f}_{(\mathfrak n_j)} (x,t,\tilde t,\tau) \\
&\sim \mathfrak Z^{(p_a),N_c,N_f,N_f-1}_{(\mathfrak n_j)} (x,t,\tilde t',\tau) \times \prod_{b = 1}^{N_f} \prod_{r = 1}^{p_b} \prod_{k = 1}^{\sum_{n = 1}^r \mathfrak n_{(b,n)}} \left(-\tilde t_{N_f}^{\frac{1}{2}} t_b^{\frac{1}{2}} \tau \upsilon^\frac{r-1}{2} x^{k-1}\right)^{\pm1}
\end{aligned}
\end{align}
as $i \tilde M_{N_f} \rightarrow \pm\infty$ where $\tilde t' = (\tilde t_1,\ldots,\tilde t_{N_f-1})$. Thus, in order to obtain a regular expression, we also need to scale $w$ such that $w \sim \tilde t_{N_f}^{\mp\frac{1}{2}}$. As a result, the left hand side of \eqref{eq:N=2 vpf duality 0} has the following large mass limits:
\begin{align}
\lim_{i \tilde M_{N_f} \rightarrow \pm\infty} Z_\text{vortex}^{(p_a),N_c,N_f,N_f}(x,t,\tilde t,\tau,w' \tilde t_{N_f}^{\mp\frac{1}{2}} \tau^{\mp\frac{1}{2}} x^{\pm\frac{1}{2}}) = Z_\text{vortex}^{(p_a),N_c,N_f,N_f-1,\pm\frac{1}{2}}(x,t,\tilde t',\tau,\mp w').
\end{align}
On the other hand, the right hand side of \eqref{eq:N=2 vpf duality 0} is a little bit complicated because there are additional factors from the singlet chiral multiplets. Firstly we observe that $Z_\text{antivortex}^{(n-p_a)}$ has similar limits:
\begin{align}
\begin{aligned}
& \lim_{i \tilde M_{N_f} \rightarrow \pm\infty} Z_\text{antivortex}^{(n-p_a),n N_f-N_c,N_f,N_f}(x,t^{-1},\tilde t^{-1},\tau^{-1} x^\delta,w'^{-1} \tilde t_{N_f}^{\pm\frac{1}{2}} \tau^{\pm\frac{1}{2}} x^{\mp\frac{1}{2}}) \\
&= Z_\text{antivortex}^{(n-p_a),n N_f-N_c,N_f,N_f-1,\pm\frac{1}{2}}(x,t^{-1},\tilde t'^{-1},\tau^{-1} x^\delta,\mp w'^{-1} x^{\mp\frac{1}{2} (2-\delta)}).
\end{aligned}
\end{align}
For the singlet chiral part, it apparently is independent of $\tilde t_{N_f}$ because $\prod_{a = 1}^{N_f} t_a = \prod_{a = 1}^{N_f} \tilde t_a = 1$. However, when we take the limits $i M_{N_f} \rightarrow \pm\infty$, we should relax that condition by shifting $\tau \rightarrow \tau \prod_{a = 1}^{N_f} t_a \tilde t_a$ because  we utilize a holonomy of $U(N_f) \sim SU(N_f) \times U(1)_A$, not that of $SU(N_f)$. Therefore, again putting $w = w' \tilde t_{N_f}^{\pm\frac{1}{2}} \tau^{\mp\frac{1}{2}} x^{\pm\frac{1}{2}}$, we has the following limit of the singlet chiral part:
\begin{align}
\lim_{i \tilde M_{N_f} \rightarrow \pm\infty} \prod_{i = 1}^{n} \frac{Z_\text{chiral}(x,w' \tilde t_{N_f}^{\mp\frac{1}{2}} \tau^{\mp\frac{1}{2}} x^{\pm\frac{1}{2}} \mathfrak t^{-\frac{1}{2}} \tau^{-N_f} x^{\Delta_i})}{Z_\text{chiral}(x,w' \tilde t_{N_f}^{\mp\frac{1}{2}} \tau^{\mp\frac{1}{2}} x^{\pm\frac{1}{2}} \mathfrak t^{\frac{1}{2}} \tau^{N_f} x^{2-\Delta_i})} = \prod_{i = 1}^{n} \frac{Z_\text{chiral}(x,w' \mathfrak t'^{-\frac{1}{2}} \tau^{-N_f+\frac{1}{2}} x^{\Delta_i-\frac{1}{2}})^{\delta_{\pm,-}}}{Z_\text{chiral}(x,w' \mathfrak t'^{-\frac{1}{2}} \tau^{N_f-\frac{1}{2}} x^{2-\Delta_i+\frac{1}{2}})^{\delta_{\pm,+}}}
\end{align}
where $\mathfrak t = \prod_{a = 1}^{N_f} t_a \tilde t_a$ and $\mathfrak t' = \mathfrak t/\tilde t_{N_f}$. $\delta_{\pm,-}$ and $\delta_{\pm,+}$ are the Kronecker delta symbol and not related to $\delta = \frac{2}{n+1}$. Combining the results, we obtain the following relation for $N_f-1$ antifundamental matters:
\begin{align}
\begin{aligned}
&\quad Z_\text{vortex}^{(p_a),N_c,N_f,N_f-1,\pm\frac{1}{2}}(x,t,\tilde t,\tau,\mp w) \\
&= Z_\text{antivortex}^{(n-p_a),n N_f-N_c,N_f,N_f-1,\pm\frac{1}{2}}(x,t^{-1},\tilde t^{-1},\tau^{-1} x^\delta,\mp w^{-1} x^{\mp\frac{1}{2} (2-\delta)}) \times \prod_{i = 1}^{n} \frac{Z_\text{chiral}(x,w \tau^{-N_f+\frac{1}{2}} x^{\Delta_i-\frac{1}{2}})^{\delta_{\pm,-}}}{Z_\text{chiral}(x,w \tau^{N_f-\frac{1}{2}} x^{2-\Delta_i+\frac{1}{2}})^{\delta_{\pm,+}}}.
\end{aligned}
\end{align}
Furthermore, repeating this procedure, we finally obtain a relation for $N_a$ antifundamental matters as follows:
\begin{align}
\begin{aligned} \label{eq:N=2 vpf duality}
&\quad Z_\text{vortex}^{(p_a),N_c,N_f,N_a,\kappa}(x,t,\tilde t,\tau,\mathfrak w) \\
&= Z_\text{antivortex}^{(n-p_a),n N_f-N_c,N_f,N_a,\kappa}(x,t^{-1},\tilde t^{-1},\tau^{-1} x^\delta,\mathfrak w^{-1} x^{-\kappa (2-\delta)}) \times \prod_{i = 1}^{n} \frac{Z_\text{chiral}(x,w \tau^{-\frac{N_f+N_a}{2}} x^{\tilde \Delta_i})^{\delta_{N_f-N_a,-2 \kappa}}}{Z_\text{chiral}(x,w \tau^{\frac{N_f+N_a}{2}} x^{2-\tilde \Delta_i})^{\delta_{N_f-N_a,2 \kappa}}}
\end{aligned}
\end{align}
where $\mathfrak w = (-1)^{-\kappa-\frac{N_f-N_a}{2}} w$ and $\tilde \Delta_i = \frac{N_f+N_a}{2}-\frac{2}{n+1} (N_c-1-i)$. Note that only the the CS coupling $\kappa$ satisfying $|\kappa| \leq \frac{N_f-N_a}{2}$ is obtained in this way. The left hand side is the vortex partition function for a theory with $N_a$ antifundamental matters. The right hand side is the vortex partition function for a $U(n N_f-N_c)_\kappa$ theory with $N_f$ fundamental, $N_a$ antifundamental, one adjoint and additional singlet matters, or equivalently $U(n N_f-N_c)_{-\kappa}$ theory with $N_a$ fundamental, $N_f$ antifundamental, one adjoint and additional singlet matters. The appearance of additional singlet matters depends on the values of $\kappa\pm\frac{N_f-N_a}{2}$.

For \eqref{eq:N=2 avpf duality 0}, exactly the same thing happens if we change $x,t,\tilde t,\tau,w \rightarrow x^{-1},t^{-1},\tilde t^{-1},\tau^{-1},w^{-1}$. Thus, we have
\begin{align}
\begin{aligned} \label{eq:N=2 avpf duality}
&\quad Z_\text{antivortex}^{(p_a),N_c,N_f,N_a,\kappa}(x,t,\tilde t,\tau,\mathfrak w) \\
&= Z_\text{vortex}^{(n-p_a),n N_f-N_c,N_f,N_a,\kappa}(x,t^{-1},\tilde t^{-1},\tau^{-1} x^\delta,\mathfrak w^{-1} x^{-\kappa (2-\delta)}) \times \prod_{i = 1}^{n} \frac{Z_\text{chiral}(x,w^{-1} \tau^{-\frac{N_f+N_a}{2}} x^{\tilde \Delta_i})^{\delta_{N_f-N_a,2 \kappa}}}{Z_\text{chiral}(x,w^{-1} \tau^{\frac{N_f+N_a}{2}} x^{2-\tilde \Delta_i})^{\delta_{N_f-N_a,-2 \kappa}}}.
\end{aligned}
\end{align}
For \eqref{eq:N=2 vpf duality} and \eqref{eq:N=2 avpf duality}, it is important to note that $\mathfrak w^{-1} x^{-\kappa(2-\delta)}$ appears on the right hand side instead of $\mathfrak w^{-1}$. The extra factor $x^{-\kappa (2-\delta)}$ indicates the presence of the mixed CS term \eqref{eq:CS RG}. Indeed, the identities \eqref{eq:N=2 pert duality}, \eqref{eq:N=2 vpf duality} and \eqref{eq:N=2 avpf duality} together imply the index agreement for the duality we propose for chiral-like theories, which is nontrivial evidence of the proposed duality. We carry out heavy numerical checks for the identities 
\eqref{eq:N=2 vpf duality} and \eqref{eq:N=2 avpf duality}.

\subsection{Examples}
Let us consider some interesting examples.

\paragraph{$N_c = n N_f$ case}
The first example is $N_c = n N_f$ case. In that case, since the dual gauge group is absent, the dual theory is only described by singlet matters. Those singlet matters would be coupled via a nontrivial superpotential. However,  when singlet matters have unitarity violating $R$-charges, additional IR symmetries would emerge and correct their unitarity violating $R$-charges. Then the superpotential becomes irrelevant such that the theory flows to the free theory in IR. For example, let us consider the $U(2)$ theory with $N_f = N_a = 1$ flavor $Q,\tilde Q$, one adjoint $X$ and the superpotential $W = X^3$, which was examined in \cite{Kim:2013cma}. Its dual theory is described by singlet matters $V_{0,\pm},V_{1,\pm},M_0,M_1$ with the superpotential
\begin{align}
\begin{aligned}
W &\sim V_{1,+} V_{1,-} M_0+V_{1,+} V_{0,-} M_1+V_{0,+} V_{1,-} M_1 \\
&\quad +V_{0,+} V_{0,-} M_0 \left(V_{1,+} V_{0,-} M_0+V_{0,+} V_{1,-} M_0+V_{0,+} V_{0,-} M_1\right) \\
&\quad +\left(V_{0,+} V_{0,-} M_0\right)^3.
\end{aligned}
\end{align}
Preserving that superpotential, those six matters cannot have $R$-charges larger than or equal to 1/2 simultaneously, which is however required due to the unitarity. In fact, we expect that in IR new symmetries emerge and mix with the $R$-symmetry such that their IR $R$-charges become 1/2. Then those matters must be free and accordingly the superpotential is also irrelevant.

There are many other examples that the dual gauge group is absent and some singlet matters have even negative $R$-charges. For example, the $U(n N_f)_0$ theory with $N_f = N_a \geq \frac{2}{n-1}$ flavors has monopole operators of zero or negative $R$-charges. Even in that case, one can check the index agreement using the factorized index we obtained. Indeed, we have checked many of such cases for several values of $N_f$ and $n$. Thus, we expect, in those cases, the theories  flow to free theories in IR.

\paragraph{$N_c = 1$ case}
Another interesting example is $N_c = 1$ case. In that case, the original theory is an abelian theory while the dual theory is the $U(n N_f-1)_{-\kappa}$ theory. However, the adjoint matter is perturbatively decoupled from the abelian gauge theory. If we assume there is no nonperturbative effect coupling the adjoint matter to the gauge interacting sector, the theory is decomposed into two sectors decoupled from each other: the gauge interacting sector and the singlet matter sector with the superpotential $W = X^{n+1}$. Interestingly the gauge interacting sector has another Seiberg-like dual description with the gauge group $U(N_f-1)_{-\kappa}$ \cite{Benini:2011mf}, which is also known as the Aharony duality if $\kappa = N_f-N_a = 0$ \cite{Aharony:1997gp}. Therefore, we have three different UV theories flowing to the same IR fixed point:
\begin{itemize}
\item $U(1)_\kappa$ gauge theory with $N_f$ fundamental and $N_a$ antifundamental matters + a decoupled matter $X$ with the superpotential $W = X^{n+1}$.
\item $U(n N_f-1)_{-\kappa}$ gauge theory with $N_a$ fundamental, $N_f$ antifundamental, one adjoint matters; additional singlet matters and a superpotential determined by (\ref{eq:superpotential 1}-\ref{eq:superpotential 3}).
\item $U(N_f-1)_{-\kappa}$ gauge theory with $N_a$ fundamental, $N_f$ antifundamental matters; additional singlet matters and a superpotential determined by (\ref{eq:superpotential 1}-\ref{eq:superpotential 3}) with $n = 1$ with a decoupled matter $X$ with the superpotential $W = X^{n+1}$.
\end{itemize}
We have checked for several values of $\kappa,N_f,N_a$ and $n$ that the three indices of them coincide.

\appendix
\section{Detailed computations for factorization} \label{sec:detailed computation}
In appendix \ref{sec:detailed computation} we examine the detailed computation to obtain the factorized form of the 3d $\mathcal N = 2$ superconformal index in the presence of an adjoint matter. As explained in section \ref{sec:factorization}, the strategy is that we explicitly compute the contour integral \eqref{eq:matrix integral}
\begin{align*}
&\quad I(x,t,\tilde t,\tau,\upsilon,w) \\
&=\sum_{m\in\mathbb Z^{N_c}/S_{N_c}} \frac{1}{|\mathcal W_m|} \oint_{|z_j| = 1} \left(\prod_{j=1}^{N_c} \frac{dz_j}{2 \pi i z_j}\right) e^{-S_\text{CS}(z,m)} w^{\sum_j m_j} Z_\text{vector}(x,z,m) Z_\text{chiral}(x,t,\tilde t,\tau,\upsilon,z,m)
\end{align*}
by evaluating the residue at the following type of a pole:
\begin{align} \label{eq:intersection2}
z_j = t_{b_j}^{-1} \tau^{-1} \upsilon^{-l_j+1} x^{-\sum_{n = 0}^{l_j-1}(|m_{\mathfrak p^n(j)}-m_{\mathfrak p^{n+1}(j)}|+2 k_{\mathfrak p^n(j)})}.
\end{align}
Each pole is represented by a forest graph which determines $b_j,l_j,\mathfrak p(j)$ and $k_j$. The detailed rules are explained in section \ref{sec:factorization}. Now recall the 1-loop determinant contributions (\ref{eq:vector}-\ref{eq:anti}) of the vector multiplet and the various chiral multiplets:
\begin{align*}
&\quad Z_\text{vector}(x,z,m) = \prod_{i < j}^{N_c} x^{-(m_i-m_j)} \left(1-z_i z_j^{-1} x^{m_i-m_j}\right) \left(1-z_j z_i^{-1} x^{m_i-m_j}\right), \\
&\begin{aligned}
&\quad Z_{X}(x,\upsilon,z,m) \\
&= \left(\frac{\left(\upsilon^{-1} x^2;x^2\right)_\infty}{\left(\upsilon;x^2\right)_\infty}\right)^{N_c} \prod_{i < j}^{N_c} \left(x^{-1} \upsilon\right)^{-(m_i-m_j)} \frac{\left(z_i^{-1} z_j \upsilon^{-1} x^{2+m_i-m_j};x^2\right)_\infty}{\left(z_i z_j^{-1} \upsilon x^{m_i-m_j};x^2\right)_\infty} \frac{\left(z_j^{-1} z_i \upsilon^{-1} x^{2+m_i-m_j};x^2\right)_\infty}{\left(z_j z_i^{-1} \upsilon x^{m_i-m_j};x^2\right)_\infty},
\end{aligned} \\
&\quad Z_{Q_a}(x,t,\tau,z,m) = \prod_{j = 1}^{N_c} \left(x^{-1} (-z_j) t_a \tau\right)^{-m_j/2} \frac{\left(z_j^{-1} t_a^{-1} \tau^{-1} x^{2+m_j};x^2\right)_\infty}{\left(z_j t_a \tau x^{m_j};x^2\right)_\infty}, \\
&\quad Z_{\tilde Q^{\tilde b}}(x,\tilde t,\tau,z,m) = \prod_{j = 1}^{N_c} \left(x^{-1} (-z_j)^{-1} \tilde t_a \tau\right)^{-m_j/2} \frac{\left(z_j \tilde t_a^{-1} \tau^{-1} x^{2+m_j};x^2\right)_\infty}{\left(z_j^{-1} \tilde t_a \tau x^{m_j};x^2\right)_\infty}
\end{align*}
where $X,Q_a,\tilde Q^{\tilde b}$ denote the adjoint, fundamental and antifundamental matters. At the pole \eqref{eq:intersection2}, those 1-loop contributions have the following values:
\begin{align}
\begin{aligned} \label{eq:vector2}
&\quad Z_\text{vector}(x,z,m) \\
&= \prod_{i < j}^{N_c} x^{-\sum_{n = 0}^{l_i-1} \mathfrak m_i^n+\sum_{n = 0}^{l_j-1} \mathfrak m_j^n} \\
&\qquad \times \left(1-t_{b_i}^{-1} t_{b_j} \upsilon^{-l_i+l_j} x^{-\sum_{n = 0}^{l_i-1} (|\mathfrak m_i^n|-\mathfrak m_i^n+2 k_i^n)+\sum_{n = 0}^{l_j-1} (|\mathfrak m_j^n|-\mathfrak m_j^n+2 k_j^n)}\right) \\
&\qquad \times \left(1-t_{b_j}^{-1} t_{b_i} \upsilon^{-l_j+l_i} x^{-\sum_{n = 0}^{l_j-1} (|\mathfrak m_j^n|+\mathfrak m_j^n+2 k_j^n)+\sum_{n = 0}^{l_i-1} (|\mathfrak m_i^n|+\mathfrak m_i^n+2 k_i^n)}\right),
\end{aligned}
\end{align}
\begin{align}
\begin{aligned}
&\quad Z_{X}(x,\upsilon,z,m) \\
&= \left(\frac{\left(\upsilon^{-1} x^2;x^2\right)_\infty}{\left(\upsilon;x^2\right)_\infty}\right)^{N_c} \prod_{i < j}^{N_c} \left(x^{-1} \upsilon\right)^{-\sum_{n = 0}^{l_i-1} \mathfrak m_i^n+\sum_{n = 0}^{l_j-1} \mathfrak m_j^n} \\
&\quad \qquad \times \frac{\left(t_{b_i} t_{b_j}^{-1} \upsilon^{l_i-l_j-1} x^{2+\sum_{n = 0}^{l_i-1} (|\mathfrak m_i^n|+\mathfrak m_i^n+2 k_i^n)-\sum_{n = 0}^{l_j-1} (|\mathfrak m_j^n|+\mathfrak m_j^n+2 k_j^n)};x^2\right)_\infty}{\left(t_{b_i}^{-1} t_{b_j} \upsilon^{-l_i+l_j+1} x^{-\sum_{n = 0}^{l_i-1} (|\mathfrak m_i^n|-\mathfrak m_i^n+2 k_i^n)+\sum_{n = 0}^{l_j-1} (|\mathfrak m_j^n|-\mathfrak m_j^n+2 k_j^n)};x^2\right)'_\infty} \\
&\quad \qquad \times \frac{\left(t_{b_j} t_{b_i}^{-1} \upsilon^{l_j-l_i-1} x^{2+\sum_{n = 0}^{l_j-1} (|\mathfrak m_j^n|-\mathfrak m_j^n+2 k_j^n)-\sum_{n = 0}^{l_i-1} (|\mathfrak m_i^n|-\mathfrak m_i^n+2 k_i^n)};x^2\right)_\infty}{\left(t_{b_j}^{-1} t_{b_i} \upsilon^{-l_j+l_i+1} x^{-\sum_{n = 0}^{l_j-1} (|\mathfrak m_j^n|+\mathfrak m_j^n+2 k_j^n)+\sum_{n = 0}^{l_i-1} (|\mathfrak m_i^n|+\mathfrak m_i^n+2 k_i^n)};x^2\right)'_\infty},
\end{aligned}
\end{align}
\begin{align}
&\begin{aligned}
&\quad Z_{Q_a}(x,t,\tau,z,m) \\
&=\prod_{j = 1}^{N_c} (-1)^{-\sum_{n = 0}^{l_j-1} \mathfrak m_j^n/2} \left(x^{-1-\sum_{n = 0}^{l_j-1} (|\mathfrak m_j^n|+2 k_j^n)} t_{b_j}^{-1} t_a \upsilon^{-l_j+1}\right)^{-\sum_{n = 0}^{l_j-1} \mathfrak m_j^n/2} \\
&\qquad \times \frac{\left(t_{b_j} t_a^{-1} \upsilon^{l_j-1} x^{2+\sum_{n = 0}^{l_j-1} (|\mathfrak m_j^n|+\mathfrak m_j^n+2 k_j^n)};x^2\right)_\infty}{\left(t_{b_j}^{-1} t_a \upsilon^{-l_j+1} x^{-\sum_{n = 0}^{l_j-1} (|\mathfrak m_j^n|-\mathfrak m_j^n+2 k_j^n)};x^2\right)'_\infty},
\end{aligned} \\
&\begin{aligned}
&\quad Z_{\tilde Q^{\tilde b}}(x,\tilde t,\tau,z,m) \\
&= \prod_{j = 1}^{N_c} (-1)^{\sum_{n = 0}^{l_j-1} \mathfrak m_j^n/2} \left(x^{-1+\sum_{n = 0}^{l_j-1} (|\mathfrak m_j^n|+2 k_j^n)} t_{b_j} \tilde t_a \tau^2 \upsilon^{l_j-1}\right)^{-\sum_{n = 0}^{l_j-1} \mathfrak m_j^n/2} \\
&\qquad \times \frac{\left(t_{b_j}^{-1} \tilde t_a^{-1} \tau^{-2} \upsilon^{-l_j+1} x^{2-\sum_{n = 0}^{l_j-1} (|\mathfrak m_j^n|-\mathfrak m_j^n+2 k_j^n)};x^2\right)_\infty}{\left(t_{b_j} \tilde t_a \tau^2 \upsilon^{l_j-1} x^{\sum_{n = 0}^{l_j-1} (|\mathfrak m_j^n|+\mathfrak m_j^n+2 k_j^n)};x^2\right)_\infty}
\end{aligned}
\end{align}
where we have defined $\mathfrak m_j = m_j-m_{\mathfrak p(j)}$ and use shorthand notations $\mathfrak m_j^n = \mathfrak m_{\mathfrak p^n(j)}$ and $k_j^n = k_{\mathfrak p^n(j)}$. The prime symbol indicates that the zero factor in the $q$-Pochhammer symbol is omitted. Now $\mathfrak m_j$ and $k_j$ always appear as a combination of $(|\mathfrak m_j|+\mathfrak m_j)/2+k_j$ and $(|\mathfrak m_j|-\mathfrak m_j)/2+k_j$. Thus, we define $\mathfrak n_j = (|\mathfrak m_j|+\mathfrak m_j)/2+k_j$ and $\bar{\mathfrak n}_j = (|\mathfrak m_j|-\mathfrak m_j)/2+k_j$, which will be interpreted as vortex charges.

Let us have a look at the 1-loop contribution of the vector multiplet \eqref{eq:vector2}. The first monomial factor is written in terms of $\mathfrak n_j$ and $\bar{\mathfrak n}_j$ in a simple way as follows:
\begin{align}
x^{-\sum_{n = 0}^{l_i-1} \mathfrak m_i+\sum_{n = 0}^{l_j-1} \mathfrak m_j} = x^{-\sum_{n = 0}^{l_i-1} (\mathfrak n_i^n-\bar{\mathfrak n}_i^n)+\sum_{n = 0}^{l_j-1} (\mathfrak n_j^n-\bar{\mathfrak n}_j^n)}
\end{align}
where $\mathfrak n_j^n = \mathfrak n_{\mathfrak p^n(j)}$ and $\bar{\mathfrak n}_j^n = \bar{\mathfrak n}_{\mathfrak p^n(j)}$.
The other factors are also written in terms of $\mathfrak n_j,\bar{\mathfrak n}_j$. Note that the second factor is independent of $\mathfrak n_j$ while the third factor is independent of $\bar{\mathfrak n}_j$. Furthermore, we can decompose them into various factors to extract perturbative contributions, which are independent of $\mathfrak n_j,\bar{\mathfrak n}_j$. For example, the third line is decomposed as follows:
\begin{align}
\begin{aligned}
&\quad \left(1-t_{b_j}^{-1} t_{b_i} \upsilon^{-l_j+l_i} x^{-2 \sum_{n = 0}^{l_j-1} \mathfrak n_j^n+2 \sum_{n = 0}^{l_i-1} \mathfrak n_i^n}\right) \\
&= \frac{\left(t_{b_j}^{-1} t_{b_i} \upsilon^{-l_j+l_i} x^{-2 \sum_{n = 0}^{l_j-1} \mathfrak n_j^n+2 \sum_{n = 0}^{l_i-1} \mathfrak n_i^n};x^2\right)_\infty}{\left(t_{b_j}^{-1} t_{b_i} \upsilon^{-l_j+l_i} x^{-2 \sum_{n = 0}^{l_j-1} \mathfrak n_j^n+2 \sum_{n = 0}^{l_i-1} \mathfrak n_i^n+2};x^2\right)_\infty} \\
&= \left(-t_{b_j}^{-1/2} t_{b_i}^{1/2} \upsilon^{-(l_j-l_i)/2} x^{-\sum_{n = 0}^{l_j-1} \mathfrak n_j^n+\sum_{n = 0}^{l_i-1} \mathfrak n_i^n}\right) \left(2 \sinh \frac{i M_{b_i}-i M_{b_j}+i \nu (l_i-l_j)}{2}\right) \\
&\quad \times \left(\prod_{k = 1}^{\sum_{n = 0}^{l_j-1} \mathfrak n_j^n} \frac{2 \sinh \frac{i M_{b_i}-i M_{b_j}+i \nu (l_i-l_j)+2 \gamma k}{2}}{2 \sinh \frac{i M_{b_i}-i M_{b_j}+i \nu (l_i-l_j)+2 \gamma (k-1-\sum_{n = 0}^{l_i-1} \mathfrak n_i^n)}{2}}\right) \\
&\quad \times \left(\prod_{k = 1}^{\sum_{n = 0}^{l_i-1} \mathfrak n_i^n} \frac{2 \sinh \frac{i M_{b_j}-i M_{b_i}+i \nu (l_j-l_i)+2 \gamma k}{2}}{2 \sinh \frac{i M_{b_j}-i M_{b_i}+i \nu (l_j-l_i)+2 \gamma (k-1-\sum_{n = 0}^{l_j-1} \mathfrak n_j^n)}{2}}\right)
\end{aligned}
\end{align}
where we have used the following identities:\footnote{Recall $(a;q)_n=\prod_{k=0}^{n-1} (1-a q^k)$.}
\begin{align}
&\begin{aligned}
&\quad \left(t_{b_j}^{-1} t_{b_i} \upsilon^{-l_j+l_i} x^{-2 \sum_{n = 0}^{l_j-1} \mathfrak n_j^n+2 \sum_{n = 0}^{l_i-1} \mathfrak n_i^n};x^2\right)_\infty \\
&= \left(t_{b_j}^{-1} t_{b_i} \upsilon^{-l_j+l_i} ;x^2\right)_\infty \\
&\quad \times \left(\prod_{k = 0}^{\sum_{n = 0}^{l_j-1} \mathfrak n_j^n} -t_{b_j}^{-1} t_{b_i} \upsilon^{-l_j+l_i} x^{-2-2 k}\right) \frac{\left(t_{b_j} t_{b_i}^{-1} \upsilon^{l_j-l_i} x^2;x^2\right)_{\sum_{n = 0}^{l_j-1} \mathfrak n_j^n}}{\left(t_{b_j}^{-1} t_{b_i} \upsilon^{-l_j+l_i} x^{-2 \sum_{n = 0}^{l_j-1} \mathfrak n_j^n};x^2\right)_{\sum_{n = 0}^{l_i-1} \mathfrak n_i^n}},
\end{aligned} \\
&\begin{aligned}
&\quad \left(t_{b_j}^{-1} t_{b_i} \upsilon^{-l_j+l_i} x^{-2 \sum_{n = 0}^{l_j-1} \mathfrak n_j^n+2 \sum_{n = 0}^{l_i-1} \mathfrak n_i^n+2};x^2\right)_\infty \\
&= \left(t_{b_j}^{-1} t_{b_i} \upsilon^{-l_j+l_i} x^2;x^2\right)_\infty \\
&\quad \times \left(\prod_{k = 0}^{\sum_{n = 0}^{l_j-1} \mathfrak n_j^n} -t_{b_j}^{-1} t_{b_i} \upsilon^{-l_j+l_i} x^{2 \sum_{n = 0}^{l_i-1} \mathfrak n_i^n-2 k}\right) \frac{\left(t_{b_j} t_{b_i}^{-1} \upsilon^{l_j-l_i} x^{-2 \sum_{n = 0}^{l_i-1} \mathfrak n_i^n};x^2\right)_{\sum_{n = 0}^{l_j-1} \mathfrak n_j^n}}{\left(t_{b_j}^{-1} t_{b_i} \upsilon^{-l_j+l_i} x^2;x^2\right)_{\sum_{n = 0}^{l_i-1} \mathfrak n_i^n}}.
\end{aligned}
\end{align}
The second line of \eqref{eq:vector2} gives the same expression but $i$ and $j$ are interchanged and $\mathfrak n_j$ is replaced by $\bar{\mathfrak n}_j$. As a result the vector multiplet contribution $Z_\text{vector}$ can be written as
\begin{align}
\begin{aligned} \label{eq:fact vector}
&\quad Z_\text{vector}(x,z,m) \\
&= \prod_{\substack{i,j = 1 \\ (i \neq j)}}^{N_c} \left(2 \sinh \frac{i M_{b_i}-i M_{b_j}+i \nu (l_i-l_j)}{2}\right) \\
&\qquad \times \left(\prod_{k = 1}^{\sum_{n = 0}^{l_j-1} \mathfrak n_j^n} \frac{\sinh \frac{i M_{b_i}-i M_{b_j}+i \nu (l_i-l_j)+2 \gamma k}{2}}{\sinh \frac{i M_{b_i}-i M_{b_j}+i \nu (l_i-l_j)+2 \gamma (k-1-\sum_{n = 0}^{l_i-1} \mathfrak n_i^n)}{2}}\right) \\
&\qquad \times \left(\prod_{k = 1}^{\sum_{n = 0}^{l_j-1} \bar{\mathfrak n}_j^n} \frac{\sinh \frac{i M_{b_i}-i M_{b_j}+i \nu (l_i-l_j)+2 \gamma k}{2}}{\sinh \frac{i M_{b_i}-i M_{b_j}+i \nu (l_i-l_j)+\gamma (k-1-\sum_{n = 0}^{l_i-1} \bar{\mathfrak n}_i^n)}{2}}\right).
\end{aligned}
\end{align}
One can see that it consists of three parts: the $\mathfrak n_j,\bar{\mathfrak n}_j$-independent part, the $\mathfrak n_j$-dependent part and the $\bar{\mathfrak n}_j^n$-dependent part. The other 1-loop contributions can be written in the same manner. The adjoint matter contribution $Z_X$ is given by
\begin{align}
\begin{aligned} \label{eq:fact adjoint}
&\quad Z_X (x,\upsilon,z,m) \\
&= \left(\frac{\left(\upsilon^{-1} x^2;x^2\right)_\infty}{\left(\upsilon;x^2\right)_\infty}\right)^{N_c} \prod_{\substack{i,j = 1 \\ (i \neq j)}}^{N_c} \frac{\left(t_{b_i} t_{b_j}^{-1} \upsilon^{l_i-l_j-1} x^2;x^2\right)_\infty}{\left(t_{b_j}^{-1} t_{b_i} \upsilon^{-l_j+l_i+1};x^2\right)'_\infty} \\
&\quad \qquad \times \left(\prod_{k = 1}^{\sum_{n = 0}^{l_j-1} \mathfrak n_j^n} \frac{\sinh \frac{i M_{b_i}-i M_{b_j}+i \nu (l_i-l_j-1)+2 \gamma (k-1-\sum_{n = 0}^{l_i-1} \mathfrak n_i^n)}{2}}{\sinh \frac{i M_{b_i}-i M_{b_j}+i \nu (l_i-l_j+1)+2 \gamma k}{2}}\right) \\
&\quad \qquad \times \left(\prod_{k = 1}^{\sum_{n = 0}^{l_j-1} \bar{\mathfrak n}_j^n} \frac{\sinh \frac{i M_{b_i}-i M_{b_j}+i \nu (l_i-l_j-1)+2 \gamma (k-1-\sum_{n = 0}^{l_i-1} \bar{\mathfrak n}_i^n)}{2}}{\sinh \frac{i M_{b_i}-i M_{b_j}+i \nu (l_i-l_j+1)+2 \gamma k}{2}}\right)
\end{aligned}
\end{align}
where the following identities are used:
\begin{align}
&\begin{aligned}
&\quad \left(t_{b_i} t_{b_j}^{-1} \upsilon^{l_i-l_j-1} x^{2+2 \sum_{n = 0}^{l_i-1} \mathfrak n_i^n-2 \sum_{n = 0}^{l_j-1} \mathfrak n_j^n};x^2\right)_\infty \\
&= \left(t_{b_i} t_{b_j}^{-1} \upsilon^{l_i-l_j-1} x^2;x^2\right)_\infty \\
&\quad \times \left(\prod_{k = 0}^{\sum_{n = 0}^{l_j-1} \mathfrak n_j^n-1} -t_{b_i} t_{b_j}^{-1} \upsilon^{l_i-l_j-1} x^{2 \sum_{n = 0}^{l_i-1} \mathfrak n_i^n-2 k}\right) \frac{\left(t_{b_i}^{-1} t_{b_j} \upsilon^{-l_i+l_j+1} x^{-2 \sum_{n = 0}^{l_i-1} \mathfrak n_i^n};x^2\right)_{\sum_{n = 0}^{l_j-1} \mathfrak n_j^n}}{\left(t_{b_i} t_{b_j}^{-1} \upsilon^{l_i-l_j-1} x^2;x^2\right)_{\sum_{n = 0}^{l_i-1} \mathfrak n_i^n}},
\end{aligned} \\
&\begin{aligned}
&\quad \left(t_{b_i}^{-1} t_{b_j} \upsilon^{-l_i+l_j+1} x^{-2 \sum_{n = 0}^{l_i-1} \bar{\mathfrak n}_i^n+2 \sum_{n = 0}^{l_j-1} \bar{\mathfrak n}_j^n};x^2\right)'_\infty \\
&= \left(t_{b_i}^{-1} t_{b_j} \upsilon^{-l_i+l_j+1};x^2\right)'_\infty \\
&\quad \times \left(\prod_{k = 0}^{\sum_{n = 0}^{l_i-1} \bar{\mathfrak n}_i^n-1} -t_{b_i}^{-1} t_{b_j} \upsilon^{-l_i+l_j+1} x^{-2-2 k}\right) \frac{\left(t_{b_i} t_{b_j}^{-1} \upsilon^{l_i-l_j-1} x^2;x^2\right)_{\sum_{n = 0}^{l_i-1} \bar{\mathfrak n}_i^n}}{\left(t_{b_i}^{-1} t_{b_j} \upsilon^{-l_i+l_j+1} x^{-2 \sum_{n = 0}^{l_i-1} \bar{\mathfrak n}_i^n};x^2\right)_{\sum_{n = 0}^{l_j-1} \bar{\mathfrak n}_j^n}},
\end{aligned}
\end{align}
\begin{align}
&\begin{aligned}
&\quad \left(t_{b_j} t_{b_i}^{-1} \upsilon^{l_j-l_i-1} x^{2+2 \sum_{n = 0}^{l_j-1} \bar{\mathfrak n}_j^n-2 \sum_{n = 0}^{l_i-1} \bar{\mathfrak n}_i^n};x^2\right)_\infty \\
&= \left(t_{b_j} t_{b_i}^{-1} \upsilon^{l_j-l_i-1} x^2;x^2\right)_\infty \\
&\quad \times \left(\prod_{k = 0}^{\sum_{n = 0}^{l_i-1} \bar{\mathfrak n}_i^n-1} -t_{b_j} t_{b_i}^{-1} \upsilon^{l_j-l_i-1} x^{2 \sum_{n = 0}^{l_j-1} \bar{\mathfrak n}_j^n-2 k}\right) \frac{\left(t_{b_j}^{-1} t_{b_i} \upsilon^{-l_j+l_i+1} x^{-2 \sum_{n = 0}^{l_j-1} \bar{\mathfrak n}_j^n};x^2\right)_{\sum_{n = 0}^{l_i-1} \bar{\mathfrak n}_i^n}}{\left(t_{b_j} t_{b_i}^{-1} \upsilon^{l_j-l_i-1} x^2;x^2\right)_{\sum_{n = 0}^{l_j-1} \bar{\mathfrak n}_j^n}},
\end{aligned} \\
&\begin{aligned}
&\quad \left(t_{b_j}^{-1} t_{b_i} \upsilon^{-l_j+l_i+1} x^{-2 \sum_{n = 0}^{l_j-1} \mathfrak n_j^n+2 \sum_{n = 0}^{l_i-1} \mathfrak n_i^n};x^2\right)'_\infty \\
&= \left(t_{b_j}^{-1} t_{b_i} \upsilon^{-l_j+l_i+1};x^2\right)'_\infty \\
&\quad \times \left(\prod_{k = 0}^{\sum_{n = 0}^{l_j-1} \mathfrak n_j^n-1} -t_{b_j}^{-1} t_{b_i} \upsilon^{-l_j+l_i+1} x^{-2-2 k}\right) \frac{\left(t_{b_j} t_{b_i}^{-1} \upsilon^{l_j-l_i-1} x^2;x^2\right)_{\sum_{n = 0}^{l_j-1} \mathfrak n_j^n}}{\left(t_{b_j}^{-1} t_{b_i} \upsilon^{-l_j+l_i+1} x^{-2 \sum_{n = 0}^{l_j-1} \mathfrak n_j^n};x^2\right)_{\sum_{n = 0}^{l_i-1} \mathfrak n_i^n}}.
\end{aligned}
\end{align}
Similarly the fundamental and the antifundamental matter contributions $Z_{Q_a}$ and $Z_{\tilde Q^{\tilde b}}$ are given by
\begin{align}
&\begin{aligned} \label{eq:fact fund}
&\quad Z_{Q_a}(x,t,\tau,z,m) \\
&= \prod_{j = 1}^{N_c} (-1)^{-\sum_{n = 0}^{l_j-1} \left(\mathfrak n_j^n-\bar{\mathfrak n}_j^n\right)/2} \frac{\left(t_{b_j} t_a^{-1} \upsilon^{l_j-1} x^2;x^2\right)_\infty}{\left(t_{b_j}^{-1} t_a \upsilon^{-l_j+1};x^2\right)'_\infty} \\
&\qquad \times \left(\prod_{k = 1}^{\sum_{n = 0}^{l_j-1} \mathfrak n_j^n} 2 \sinh \frac{i M_a-i M_{b_j}-i \nu (l_j-1)+2 \gamma k}{2}\right)^{-1} \\
&\qquad \times \left(\prod_{k = 1}^{\sum_{n = 0}^{l_j-1} \bar{\mathfrak n}_j^n} 2 \sinh \frac{-i M_a+i M_{b_j}+i \nu (l_j-1)+- \gamma k}{2}\right)^{-1},
\end{aligned}
\end{align}
\begin{align}
&\begin{aligned} \label{eq:fact anti}
&\quad Z_{\tilde Q^{\tilde b}}(x,\tilde t,\tau,z,m) \\
&= \prod_{j = 1}^{N_c} (-1)^{\sum_{n = 0}^{l_j-1} \left(\mathfrak n_j^n-\bar{\mathfrak n}_j^n\right)/2} \frac{\left(t_{b_j}^{-1} \tilde t_a^{-1} \tau^{-2} \upsilon^{-l_j+1} x^2;x^2\right)_\infty}{\left(t_{b_j} \tilde t_a \tau^2 \upsilon^{l_j-1};x^2\right)_\infty} \\
&\qquad \times \left(\prod_{k = 1}^{\sum_{n = 0}^{l_j-1} \mathfrak n_j^n} 2 \sinh \frac{-i \tilde M_a-i M_{b_j}-2 i \mu-i \nu (l_j-1)+2 \gamma (k-1)}{2}\right) \\
&\qquad \times \left(\prod_{k = 1}^{\sum_{n = 0}^{l_j-1} \bar{\mathfrak n}_j^n} 2 \sinh \frac{i \tilde M_a+i M_{b_j}+2 i \mu+i \nu (l_j-1)-2 \gamma (k-1)}{2}\right)
\end{aligned}
\end{align}
where the following identities are used:
\begin{align}
&\begin{aligned}
&\quad \left(t_{b_j} t_a^{-1} \upsilon^{l_j-1} x^{2+2 \sum_{n = 0}^{l_j-1} \mathfrak n_j^n};x^2\right)_\infty \\
&= \left(t_{b_j} t_a^{-1} \upsilon^{l_j-1} x^2;x^2\right)_\infty \times \frac{1}{\left(t_{b_j} t_a^{-1} \upsilon^{l_j-1} x^2;x^2\right)_{\sum_{n = 0}^{l_j-1} \mathfrak n_j^n}},
\end{aligned} \\
&\begin{aligned}
&\quad \left(t_{b_j}^{-1} t_a \upsilon^{-l_j+1} x^{-2 \sum_{n = 0}^{l_j-1} \bar{\mathfrak n}_j^n};x^2\right)'_\infty \\
&= \left(t_{b_j}^{-1} t_a \upsilon^{-l_j+1};x^2\right)'_\infty \times \left(\prod_{k = 0}^{\sum_{n = 0}^{l_j-1} \bar{\mathfrak n}_j^n-1} -t_{b_j}^{-1} t_a \upsilon^{-l_j+1} x^{-2-2 k}\right) \left(t_{b_j} t_a^{-1} \upsilon^{l_j-1} x^2;x^2\right)_{\sum_{n = 0}^{l_j-1} \bar{\mathfrak n}_j^n},
\end{aligned}
\end{align}
\begin{align}
&\begin{aligned}
&\quad \left(t_{b_j}^{-1} \tilde t_a^{-1} \tau^{-2} \upsilon^{-l_j+1} x^{2-2 \sum_{n = 0}^{l_j-1} \bar{\mathfrak n}_j^n};x^2\right)_\infty \\
&= \left(t_{b_j}^{-1} \tilde t_a^{-1} \tau^{-2} \upsilon^{-l_j+1} x^2;x^2\right)_\infty \times \left(\prod_{k = 0}^{\sum_{n = 0}^{l_j-1} \bar{\mathfrak n}_j^n-1} -t_{b_j}^{-1} \tilde t_a^{-1} \tau^{-2} \upsilon^{-l_j+1} x^{-2 k}\right) \left(t_{b_j} \tilde t_a \tau^2 \upsilon^{l_j-1};x^2\right)_{\sum_{n = 0}^{l_j-1} \bar{\mathfrak n}_j^n},
\end{aligned} \\
&\begin{aligned}
&\quad \left(t_{b_j} \tilde t_a \tau^2 \upsilon^{l_j-1} x^{2 \sum_{n = 0}^{l_j-1} \mathfrak n_j^n};x^2\right)_\infty \\
&= \left(t_{b_j} \tilde t_a \tau^2 \upsilon^{l_j-1};x^2\right)_\infty \times \frac{1}{\left(t_{b_j} \tilde t_a \tau^2 \upsilon^{l_j-1};x^2\right)_{\sum_{n = 0}^{l_j-1} \mathfrak n_j^n}}.
\end{aligned}
\end{align}

Those 1-loop contributions together with the classical contribution
\begin{align}
e^{-S_\text{CS} (z,m)} = \prod_{j = 1}^{N_c} (-1)^{-\kappa \sum_{n = 0}^{l_j-1} \left(\mathfrak n_j^n-\bar{\mathfrak n}_j^n\right)} \left(t_{b_j} \tau \upsilon^{l_j-1} x^{\sum_{n = 0}^{l_j-1} \left(\mathfrak n_j^n+\bar{\mathfrak n}_j^n\right)}\right)^{\kappa \sum_{n = 0}^{l_j-1} \left(\mathfrak n_j^n-\bar{\mathfrak n}_j^n\right)},
\end{align}
are summed over the poles classified by the labeled forest graphs we introduce in section \ref{sec:factorization} and over all possible monopole fluxes:
\begin{align} \label{eq:sum}
\sum_{m \in \mathbb Z^{N_c}/S_{N_c}} \frac{1}{|\mathcal W_m|} \sum_{\mathfrak f \in \mathfrak F} \leftrightarrow \frac{1}{N_c!} \sum_{m \in Z^{N_c}} \sum_{\mathfrak f \in \mathfrak F}
\end{align}
where the arrow means that they are equivalent sums. $\left|\mathcal W_m\right|$ is the Weyl group order of the residual gauge group left unbroken by the magnetic flux $m$. $\mathfrak F$ is a set of the labeled forest graphs of $N_c$ nodes in which each root node is labeled by $(j,\mathsf a,\mathsf k)$ while each non-root node is labeled by $(j,\mathsf k)$ with $j \in \{1,\ldots,N_c\},\mathsf a \in \{1,\ldots,N_f\}$ and $\mathsf k \geq 0$. We denote $\mathsf k$ assigned to the $j$-th node by $k_j$. Since we sum over all possible labelings, we can extract the $\mathsf k$ assignment for each node as explicit summations over $k_j$:
\begin{align}
\frac{1}{N_c!} \sum_{m \in Z^{N_c}} \sum_{\mathfrak f \in \mathfrak F} \leftrightarrow \frac{1}{N_c!} \sum_{m \in Z^{N_c}} \sum_{\mathfrak g \in \mathfrak G} \sum_{k_j \geq 0}
\end{align}
where $\mathfrak G$ is a set of the labeled forest graphs that the $k_j$ assignment is removed; i.e., each root node is now labeled by $(j,\mathsf a)$ while each non-root node is only labeled by $j$. Now recall $\mathfrak n_j$ and $\bar{\mathfrak n}_j$, which are defined by $\mathfrak n_j = (|\mathfrak m_j|+\mathfrak m_j)/2+k_j$ and $\bar{\mathfrak n}_j = (|\mathfrak m_j|-\mathfrak m_j)/2+k_j$ where $\mathfrak m_j = m_j-m_{\mathfrak p(j)}$. One can reorganize the magnetic flux and $k_j$ summations in terms of $\mathfrak n_j$ and $\bar{\mathfrak n}_j$ as follows:
\begin{align}
\sum_{m \in Z^{N_c}} \sum_{k_j \geq 0} \leftrightarrow \sum_{\mathfrak n_j \geq 0} \sum_{\bar{\mathfrak n}_j \geq 0}.
\end{align}
Thus, we have the following summation equivalent to \eqref{eq:sum}:
\begin{align}
\sum_{m \in \mathbb Z^{N_c}/S_{N_c}} \frac{1}{|\mathcal W_m|} \sum_{\mathfrak f \in \mathfrak F} \leftrightarrow \frac{1}{N_c!} \sum_{\mathfrak g \in \mathfrak G} \sum_{\mathfrak n_j \geq 0} \sum_{\bar{\mathfrak n}_j \geq 0}.
\end{align}

Since each residue factor consists of three parts: the $\mathfrak n_j,\bar{\mathfrak n}_j$-independent part, the $\mathfrak n_j$-depdendent part and the $\bar{\mathfrak n}_j^n$-dependent part, one can write down the index in the following factorized form:
\begin{align}
I(x,t,\tilde t,\tau,\upsilon,w) = \frac{1}{N_c!}\sum_{\mathfrak g \in \mathfrak G} I_\text{pert}^{\mathfrak g}(x,t,\tilde t,\tau,\upsilon) Z_\text{vortex}^{\mathfrak g}(x,t,\tilde t,\tau,\upsilon,\mathfrak w) Z_\text{antivortex}^{\mathfrak g}(x,t,\tilde t,\tau,\upsilon,\mathfrak w)
\end{align}
\begin{align}
\begin{aligned} \label{eq:pert}
&\quad I_\text{pert}^{\mathfrak g}(x,t = e^{i M},\tilde t,\tau,\upsilon = e^{i \nu}) \\
&= \left(\prod_{\substack{i,j = 1 \\ (i \neq j)}}^{N_c} 2 \sinh \frac{i M_{b_i}-i M_{b_j}+i \nu (l_i-l_j)}{2}\right) \left(\prod_{i,j = 1}^{N_c} \frac{\left(t_{b_i} t_{b_j}^{-1} \upsilon^{l_i-l_j-1} x^2;x^2\right)_\infty}{\left(t_{b_j}^{-1} t_{b_i} \upsilon^{-l_j+l_i+1};x^2\right)'_\infty}\right) \\
&\quad \times \left(\prod_{j = 1}^{N_c} \frac{\prod_{a = 1}^{N_f} \left(t_{b_j} t_a^{-1} \upsilon^{l_j-1} x^2;x^2\right)_\infty}{\prod_{a = 1}^{N_a} \left(t_{b_j} \tilde t_a \tau^2 \upsilon^{l_j-1};x^2\right)_\infty} \frac{\prod_{a = 1}^{N_a} \left(t_{b_j}^{-1} \tilde t_a^{-1} \tau^{-2} \upsilon^{-l_j+1} x^2;x^2\right)_\infty}{\prod_{a = 1}^{N_f} \left(t_{b_j}^{-1} t_a \upsilon^{-l_j+1};x^2\right)'_\infty}\right)
\end{aligned}
\end{align}
\begin{align}
Z_\text{vortex}^{\mathfrak g}(x,t,\tilde t,\tau,\upsilon,\mathfrak w) &= \sum_{\mathfrak n_j \geq 0} \mathfrak w^{\sum_{j = 1}^{N_c} \sum_{n = 0}^{l_j-1} \mathfrak n_j^n} \mathfrak Z^{\mathfrak g}_{(\mathfrak n_j)}(x,t,\tilde t,\tau,\upsilon), \\
Z_\text{antivortex}^{\mathfrak g}(x,t,\tilde t,\tau,\upsilon,\mathfrak w) &= \sum_{\bar{\mathfrak n}_j \geq 0} \mathfrak w^{-\sum_{j = 1}^{N_c} \sum_{n = 0}^{l_j-1} \bar{\mathfrak n}_j^n} \mathfrak Z^{\mathfrak g}_{(\bar{\mathfrak n}_j)}(x^{-1},t^{-1},\tilde t^{-1},\tau^{-1},\upsilon^{-1}),
\end{align}
\begin{gather}
\begin{aligned}
&\quad \mathfrak Z^{\mathfrak g}_{(\mathfrak n_j)}(x = e^{-\gamma},t = e^{i M},\tilde t = e^{i \tilde M},\tau = e^{i \mu},\upsilon = e^{i \nu}) \\
&= e^{-S^{\mathfrak g}_{(\mathfrak n_j)}(x,t,\tau,\upsilon)} \left(\prod_{\substack{i,j = 1 \\ (i \neq j)}}^{N_c} \prod_{k = 1}^{\sum_{n = 0}^{l_j-1} \mathfrak n_j^n} \frac{\sinh \frac{i M_{b_i}-i M_{b_j}+i \nu (l_i-l_j)+2 \gamma k}{2}}{\sinh \frac{i M_{b_i}-i M_{b_j}+i \nu (l_i-l_j)+2 \gamma (k-1-\sum_{n = 0}^{l_i-1} \mathfrak n_i^n)}{2}}\right) \\
&\quad \times \left(\prod_{\substack{i,j = 1 \\ (i \neq j)}}^{N_c} \prod_{k = 1}^{\sum_{n = 0}^{l_j-1} \mathfrak n_j^n} \frac{\sinh \frac{i M_{b_i}-i M_{b_j}+i \nu (l_i-l_j-1)+2 \gamma (k-1-\sum_{n = 0}^{l_i-1} \mathfrak n_i^n)}{2}}{\sinh \frac{i M_{b_i}-i M_{b_j}+i \nu (l_i-l_j+1)+2 \gamma k}{2}}\right) \\
&\quad \times \left(\prod_{j = 1}^{N_c} \prod_{k = 1}^{\sum_{n = 0}^{l_j-1} \mathfrak n_j^n} \frac{\prod_{a = 1}^{N_a} \sinh \frac{-i \tilde M_a-i M_{b_j}-2 i \mu-i \nu (l_j-1)+2 \gamma (k-1)}{2}}{\prod_{a = 1}^{N_f} \sinh \frac{i M_a-i M_{b_j}-i \nu (l_j-1)+2 \gamma k}{2}}\right),
\end{aligned} \\
e^{-S^{\mathfrak g}_{(\mathfrak n_j)}(x,t,\tau,\upsilon)} = \prod_{j = 1}^{N_c} \left(t_{b_j} \tau \upsilon^{l_j-1} x^{\sum_{n = 0}^{l_j-1}\mathfrak n_j^n}\right)^{\kappa \sum_{n = 0}^{l_j-1}\mathfrak n_j^n}
\end{gather}
where $\mathfrak w = (-1)^{-\kappa-\frac{N_f-N_a}{2}} w$ and $\mathfrak n_j^n = \mathfrak n_{\mathfrak p^n(j)},\bar{\mathfrak n}_j^n = \bar{\mathfrak n}_{\mathfrak p^n(j)}$. The prime symbol indicates that the zero factor in the $q$-Pochhammer symbol is omitted. $\mathfrak G$ is a set of the labeled forest graphs of $N_c$ nodes in which each root node is labeled by $(j,\mathsf a)$ while each non-root node is labeled by $j$ with $j \in \{1,\ldots,N_c\}$ and $\mathsf a \in \{1,\ldots,N_f\}$. $b_j,l_j$ and $\mathfrak p(j)$ are determined by the forest graph $\mathfrak g \in \mathfrak G$ as explained in section \ref{sec:factorization}.

Since $i,j$ are dummy variables, the above expressions are invariant under a permutation of $i,j$:
\begin{align}
i \rightarrow \sigma(i), \quad j \rightarrow \sigma(j), \qquad \sigma \in S_{N_c}
\end{align}
where $S_{N_c}$ is the symmetric group of degree $N_c$. On the other hand, this permutation of $i,j$ can be regarded as a relabeling of the nodes in $\mathfrak g$:
\begin{align} \label{eq:relabeling}
\begin{aligned}
(j,\mathsf a) &\rightarrow (\sigma(j),\mathsf a) \\
j &\rightarrow \sigma(j)
\end{aligned} \qquad
\begin{aligned}
& \text{for a root node,} \\
& \text{for a non-root node}
\end{aligned}
\end{align}
as well as a reordering of $\mathfrak n_j$ and $\bar{\mathfrak n}_j$:
\begin{align}
\mathfrak n_j \rightarrow \mathfrak n_{\sigma(j)}, \quad \bar{\mathfrak n}_j \rightarrow \bar{\mathfrak n}_{\sigma(j)}.
\end{align}
The reordering of $\mathfrak n_j$ and $\bar{\mathfrak n}_j$ doesn't affect $Z_\text{vortex}^{\mathfrak g}$ and $Z_\text{antivortex}^{\mathfrak g}$ because we sum over all possible $\mathfrak n_j$ and $\bar{\mathfrak n}_j$. Thus, we can conclude that $I_\text{pert}^{\mathfrak g} Z_\text{vortex}^{\mathfrak g} Z_\text{antivortex}^{\mathfrak g}$ is invariant under the relabeling of the nodes in $\mathfrak g$. Since $|S_{N_c}| = N_c!$, there are $N_c!$ relabelings of $\mathfrak g$ which all give the same contribution of $I_\text{pert}^{\mathfrak g} Z_\text{vortex}^{\mathfrak g} Z_\text{antivortex}^{\mathfrak g}$.

Indeed, \eqref{eq:pert} tells us that some of $\mathfrak g$'s have the vanishing residues because $I_\text{pert}^{\mathfrak g}$ vanishes if $b_i = b_j$ and $l_i = l_j$ for different $i$ and $j$. In other words, the residue vanishes unless the corresponding $\mathfrak g$ only contains one-branch trees each of which has different $b_j$.\footnote{We call a tree a one-branch tree if each node has at most one child node.}  In addition, since the relabelings \eqref{eq:relabeling} give the same contribution, we only take one canonical labeling and multiply its contribution by $N_c!$. Let us call a tree whose root node is labeled by $(j,\mathsf a)$ the $\mathsf a$-th tree. Then we fix the labeling of each node in the $\mathsf a$-th tree in the monotonically increasing way:
\begin{align} \label{eq:labeling}
j = \sum_{a = 1}^{\mathsf a-1} p_a+q
\end{align}
where $p_a$ is the height of the $a$-th tree, which is the total number of the nodes of the one-branch tree, and $q$ is the level of the node. If the $a$-th tree is absent, $p_a = 0$.
\begin{figure}[tbp]
\centering
\includegraphics[width=.9\textwidth]{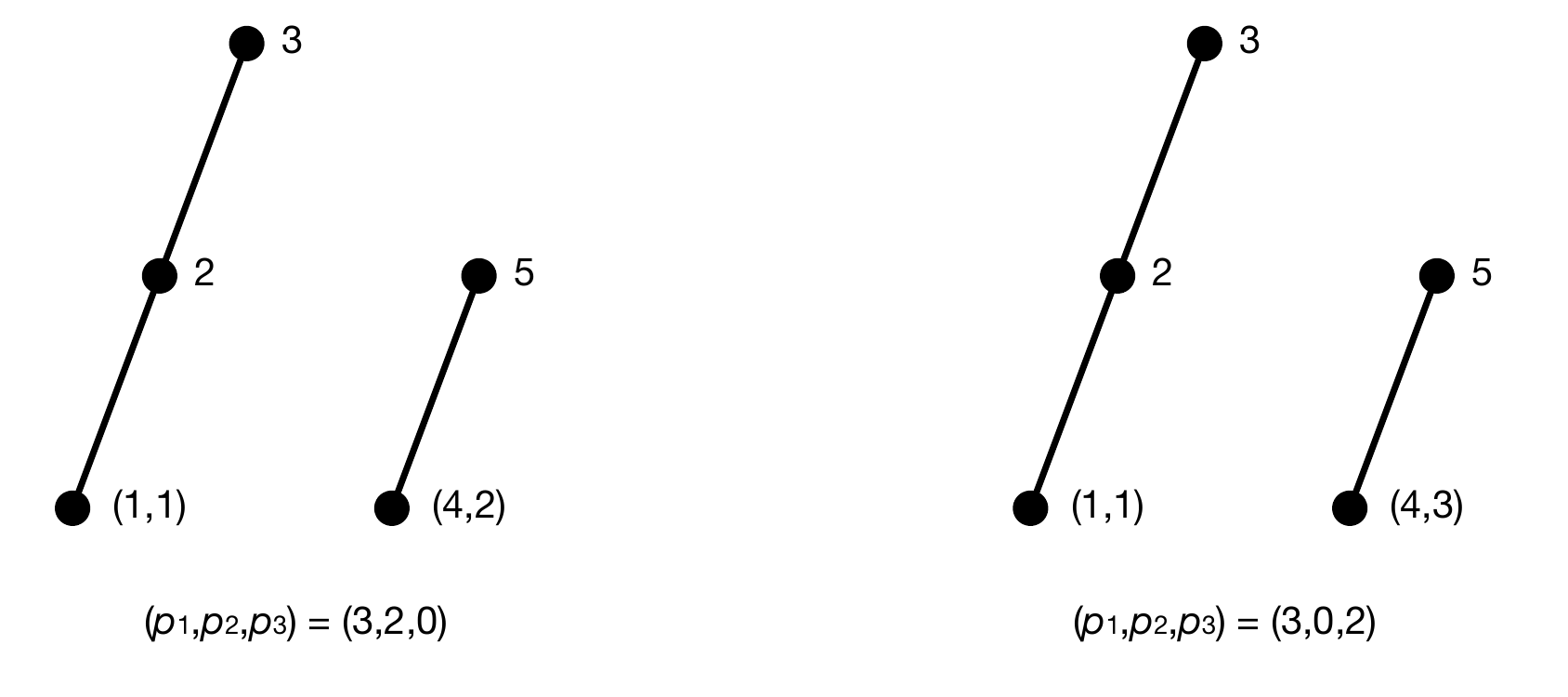}
\caption{\label{fig:one-branch} We illustrate two examples of forest graphs only containing one-branch trees for the $U(5)$ theory with three flavors. The left graph corresponds to $(p_1,p_2,p_3) = (3,2,0)$, in which the third tree is absent, while the right graph corresponds to $(p_1,p_2,p_3) = (3,0,2)$, in which the second tree is absent. Note that $p_1+p_2+p_3 = 5$. The labeling of the nodes is fixed by \eqref{eq:labeling}.}
\end{figure}
Then it is economical to label the nontrivial poles by $N_f$-tuples $(p_a) = (p_1,\ldots,p_{N_f})$ instead of the forest graphs where $p_a$'s are nonnegative integers satisfying $\sum_{a = 1}^{N_f} p_a = N_c$. Those are nothing but the one-dimensional $N_f$-colored Young diagrams of $N_c$ boxes. Therefore, one can write down $I_\text{pert}$ and $\mathfrak Z_{(n_j)}$ in terms of $(p_a)$ as follows:
\begin{align}
\begin{aligned}
&\quad I_\text{pert}^{(p_a)}(x,t = e^{i M},\tilde t,\tau,\upsilon = e^{i \nu}) \\
&= \left(\prod_{a,b = 1}^{N_f} \prod_{q = 1}^{p_a} \prod_{\substack{r = 1 \\ (\neq q \text{ if } a = b)}}^{p_b} 2 \sinh \frac{i M_{a}-i M_{b}+i \nu (q-r)}{2}\right) \left(\prod_{a,b = 1}^{N_f} \prod_{q = 1}^{p_a} \prod_{r = 1}^{p_b} \frac{\left(t_{a} t_{b}^{-1} \upsilon^{q-r-1} x^2;x^2\right)_\infty}{\left(t_{a}^{-1} t_{b} \upsilon^{-q+r+1};x^2\right)'_\infty}\right) \\
&\quad \times \left(\prod_{a = 1}^{N_f} \prod_{q = 1}^{p_a} \frac{\prod_{b = 1}^{N_f} \left(t_a t_b^{-1} \upsilon^{q-1} x^2;x^2\right)_\infty}{\prod_{b = 1}^{N_a} \left(t_a \tilde t_b \tau^2 \upsilon^{q-1};x^2\right)_\infty} \frac{\prod_{b = 1}^{N_a} \left(t_a^{-1} \tilde t_b^{-1} \tau^{-2} \upsilon^{-q+1} x^2;x^2\right)_\infty}{\prod_{b = 1}^{N_f} \left(t_a^{-1} t_b \upsilon^{-q+1};x^2\right)'_\infty}\right),
\end{aligned}
\end{align}
\begin{align}
\begin{aligned}
&\quad \mathfrak Z^{(p_a)}_{(\mathfrak n_j)}(x = e^{-\gamma},t = e^{i M},\tilde t = e^{i \tilde M},\tau = e^{i \mu},\upsilon = e^{i \nu}) \\
&= e^{-S^{(p_a)}_{(\mathfrak n_j)}(x,t,\tau,\upsilon)} \left(\prod_{a,b = 1}^{N_f} \prod_{q = 1}^{p_a} \prod_{\substack{r = 1 \\ (\neq q \text{ if } a = b)}}^{p_b} \prod_{k = 1}^{\sum_{n = 1}^r \mathfrak n_{(b,n)}} \frac{\sinh \frac{i M_a-i M_b+i \nu (q-r)+2 \gamma k}{2}}{\sinh \frac{i M_a-i M_b+i \nu (q-r)+ 2 \gamma (k-1-\sum_{n = 1}^q \mathfrak n_{(a,n)})}{2}}\right) \\
&\quad \times \left(\prod_{a,b = 1}^{N_f} \prod_{q = 1}^{p_a} \prod_{\substack{r = 1 \\ (\neq q \text{ if } a = b)}}^{p_b} \prod_{k = 1}^{\sum_{n = 1}^r \mathfrak n_{(b,n)}} \frac{\sinh \frac{i M_a-i M_b+i \nu (q-r-1)+2 \gamma (k-1-\sum_{n = 1}^q \mathfrak n_{(a,n)})}{2}}{\sinh \frac{i M_a-i M_b+i \nu (q-r+1)+2 \gamma k}{2}}\right) \\
&\quad \times \left(\prod_{b = 1}^{N_f} \prod_{r = 1}^{p_b} \prod_{k = 1}^{\sum_{n = 1}^r \mathfrak n_{(b,n)}} \frac{\prod_{a = 1}^{N_a} \sinh \frac{-i \tilde M_a-i M_b-2 i \mu-i \nu (r-1)+2 \gamma (k-1)}{2}}{\prod_{a = 1}^{N_f} \sinh \frac{i M_a-i M_b-i \nu (r-1)+2 \gamma k}{2}}\right)
\end{aligned}
\end{align}
where
\begin{align}
e^{-S^{(p_a)}_{(\mathfrak n_j)}(x,t,\tau,\upsilon)} = \prod_{b = 1}^{N_f} \prod_{r = 1}^{p_b} \left(t_b \tau \upsilon^{r-1} x^{\sum_{n = 1}^r \mathfrak n_{(b,n)}}\right)^{\kappa \sum_{n = 1}^r \mathfrak n_{(b,n)}}.
\end{align}
$\mathfrak n_{(a,q)}$ is a shorthand notation for $\mathfrak n_{q+\sum_{b = 1}^{a-1} p_b}$. We claim that this is the vortex partition function on $\mathbb R^2 \times S^1$ of the $\mathcal N = 2$ $U(N_c)_\kappa$ gauge theory with $N_f$ fundamental, $N_a$ antifundamental and one adjoint matter under the condition $|\kappa| \leq \frac{N_f-N_a}{2}$. Note that the antivortex part is obtained from the vortex part by inverting all the fugacities, $x,t,\tilde t,\tau,\upsilon,w \rightarrow x^{-1},t^{-1},\tilde t^{-1},\tau^{-1},\upsilon^{-1},\mathfrak w^{-1}$.

\section{Large mass behavior of the vortex partition function} \label{sec:large mass}
In order to prove the agreement of the vortex part for $\mathcal N = 4$ Seiberg-like duality in section \ref{sec:N=4}, we are interested in the large mass behavior of the vortex partition function of a $\mathcal N = 4$ theory. In this appendix we will choose one of the flavors and take its real mass to infinity; i.e., $i M_a \rightarrow \pm\infty$. One should note that for the vortex partition function the flavors are grouped into two distinct sets: $\{b_j\}$ and $\{b_j\}^c$. The large mass behavior of the vortex partition function depends on whether the flavor is selected from $\{b_j\}$ or from $\{b_j\}^c$. We first consider the latter case because it is rather simple. If we take the limit $i M_{\mathfrak a} \rightarrow \pm\infty$ with $\mathfrak a \in \{b_j\}^c$, we have
\begin{align}
\frac{2 \sinh \frac{i M_{\mathfrak a}-i M_{b_j}-2 i \mu+2 \gamma (k-\frac{1}{2})}{2}}{2 \sinh \frac{i M_{\mathfrak a}-i M_{b_j}+2 \gamma k}{2}} \longrightarrow \tau^{\mp1} x^{\pm\frac{1}{2}}.
\end{align}
Therefore,
\begin{align}
\begin{aligned}
&\quad \lim_{i M_{\mathfrak a} \rightarrow \pm\infty} \mathfrak Z_{(n_j)}^{\{b_j\},N_c,N_f}(x = e^{-\gamma},t = e^{i M},\tau = e^{i \mu}) \\
&= \left(\tau^{\mp1} x^{\pm\frac{1}{2}}\right)^{\sum_j n_j} \\
&\qquad \times \prod_{j=1}^{N_c} \prod_{k=1}^{n_j} \left(\prod_{i=1}^{N_c} \frac{2 \sinh \frac{i M_{b_i}-i M_{b_j}+2 i \mu+2 \gamma (k-\frac{1}{2}-n_i)}{2}}{2 \sinh \frac{i M_{b_i}-i M_{b_j}+2 \gamma (k-1-n_i)}{2}}\right) \left(\prod_{a\in \{b_j\}^c-\{\mathfrak a\}} \frac{2 \sinh \frac{i M_a-i M_{b_j}-2 i \mu+2 \gamma (k-\frac{1}{2})}{2}}{2 \sinh \frac{i M_a-i M_{b_j}+2 \gamma k}{2}}\right) \\
&= \left(\tau^{\mp1} x^{\pm\frac{1}{2}}\right)^{\sum_j n_j} \mathfrak Z_{(n_j)}^{\{b_j\},N_c,N_f-1}\left(x = e^{-\gamma},t' = e^{i M'},\tau = e^{i \mu}\right)
\end{aligned}
\end{align}
where $\mathfrak Z_{(n_j)}^{\{b_j\},N_c,N_f-1}$ contains $N_f-1$ flavors whose corresponding fugacities are given by $t'=(t_1,\ldots,t_{\mathfrak a-1},t_{\mathfrak a+1},\ldots,t_{N_f})$. From this relation we learn that in the large mass limit the vortex partition function becomes
\begin{align}
\lim_{i M_{\mathfrak a} \rightarrow \pm\infty} Z_\text{vortex}^{\{b_j\},N_c,N_f}\left(x,t,\tau,w\right) = Z_\text{vortex}^{\{b_j\},N_c,N_f-1}(x,t',\tau,w \tau^{\mp1} x^{\pm1}).
\end{align}
It shows that if we take the large mass flavor from $\{b_j\}^c$, the vortex partition function just reduces to that of $N_f-1$ flavors.

On the other hand, if we now take the $\mathfrak b$th flavor with $\mathfrak b = b_{\mathfrak j} \in \{b_j\}$ and take the limit $i M_{\mathfrak b} \rightarrow \pm\infty$, we have
\begin{align}
&\frac{2 \sinh \frac{i M_{\mathfrak b}-i M_{b_j}+2 i \mu+2 \gamma (k-\frac{1}{2}-n_i)}{2}}{2 \sinh \frac{i M_{\mathfrak b}-i M_{b_j}+2 \gamma (k-1-n_i)}{2}} \rightarrow \tau^{\pm1} x^{\mp\frac{1}{2}}, \\
&\frac{2 \sinh \frac{i M_{b_i}-i M_{\mathfrak b}+2 i \mu+2 \gamma (k-\frac{1}{2}-n_i)}{2}}{2 \sinh \frac{i M_{b_i}-i M_{\mathfrak b}+2 \gamma (k-1-n_i)}{2}} \rightarrow \tau^{\mp1} x^{\pm\frac{1}{2}}, \\
&\frac{2 \sinh \frac{i M_a-i M_{\mathfrak b}-2 i \mu+2 \gamma (k-\frac{1}{2})}{2}}{2 \sinh \frac{i M_a-i M_{\mathfrak b}+2 \gamma k}{2}} \rightarrow \tau^{\pm1} x^{\mp\frac{1}{2}}
\end{align}
where $i,j \neq \mathfrak j$ and $a \in \{b_j\}^c$. Then we also have
\begin{align}
\begin{aligned}
&\quad \lim_{i M_{\mathfrak b} \rightarrow \pm\infty} \mathfrak Z_{(n_j)}^{\{b_j\},N_c,N_f}\left(x = e^{-\gamma},t = e^{i M},\tau = e^{i \mu}\right) \\
&= \left(\tau^{\pm1} x^{\mp\frac{1}{2}}\right)^{\sum_{j \neq \mathfrak j}n_j+(N_f-2 N_c+1) n_{\mathfrak j}} \times \left(\prod_{k=1}^{n_{\mathfrak j}} \frac{2 \sinh \frac{2 i \mu+2 \gamma (k-\frac{1}{2}-n_{\mathfrak j})}{2}}{2 \sinh \frac{2 \gamma (k-1-n_{\mathfrak j})}{2}}\right) \\
&\quad \times \prod_{j=1 (\neq \mathfrak j)}^{N_c} \prod_{k=1}^{n_j} \left(\prod_{i=1 (\neq \mathfrak j)}^{N_c} \frac{2 \sinh \frac{i M_{b_i}-i M_{b_j}+2 i \mu+2 \gamma (k-\frac{1}{2}-n_i)}{2}}{2 \sinh \frac{i M_{b_i}-i M_{b_j}+2 \gamma (k-1-n_i)}{2}}\right) \left(\prod_{a\in \{b_j\}^c} \frac{2 \sinh \frac{i M_a-i M_{b_j}-2 i \mu+2 \gamma (k-\frac{1}{2})}{2}}{2 \sinh \frac{i M_a-i M_{b_j}+2 \gamma k}{2}}\right) \\
&= \left(\tau^{\pm1} x^{\mp\frac{1}{2}}\right)^{\sum_{j \neq \mathfrak j}n_j+(N_f-2 N_c+1) n_{\mathfrak j}} \times \mathfrak Z_{(n_j)}^{\{b_j\},1,1}\left(x = e^{-\gamma},1,\tau = e^{i \mu}\right) \\
&\quad \times \mathfrak Z_{(n_j)}^{\{b_j\},N_c-1,N_f-1}\left(x = e^{-\gamma},t'' = e^{i M''},\tau = e^{i \mu}\right)
\end{aligned}
\end{align}
where $t''=(t_1,\ldots,t_{\mathfrak b-1},t_{\mathfrak b+1},\ldots,t_{N_f})$. Therefore, the vortex partition function becomes
\begin{align}
\begin{aligned}\label{eq:large mass}
&\quad \lim_{i M_{\mathfrak b} \rightarrow \pm\infty} Z_\text{vortex}^{\{b_j\},N_c,N_f}(x,t,\tau,w) \\
&= Z_\text{vortex}^{1,1}(x,1,\tau,w \tau^{\mp(2 N_c-N_f-1)} x^{\pm(2 N_c-N_f-1)/2}) \times Z_\text{vortex}^{\{b_j\},N_c-1,N_f-1}(x,t'',\tau,w \tau^{\pm1} x^{\mp\frac{1}{2}}).
\end{aligned}
\end{align}
where $Z_\text{vortex}^{1,1} \equiv Z_\text{vortex}^{\{1\},1,1}$ is the vortex partition function of the $U(1)$ theory with one flavor.  Using the $q$-binomial theorem
\begin{align}
\sum_{n \geq 0} \frac{(a;q)_n}{(q;q)_n} z^n=\frac{(a z;q)_\infty}{(z;q)_\infty},
\end{align}
one can show that $Z_\text{vortex}^{1,1}$ is written as follows:
\begin{align}
Z_\text{vortex}^{1,1}(x,1,\tau,w) = \frac{\left(\tau w x^\frac{3}{2};x^2\right)_\infty}{\left(\tau^{-1} w x^\frac{1}{2};x^2\right)_\infty}.
\end{align}
Note that the right hand appears as a part of the superconformal index of a free twisted hypermultiplet. If we call the right hand side $Z_\text{hyper}$, the index of the twisted hypermultiplet can be written as
\begin{align}
I_\text{hyper}(x,\tau,w) = Z_\text{hyper}(x,\tau,w) \times Z_\text{hyper}(x^{-1},\tau^{-1},w^{-1})
\end{align}
where $Z_\text{hyper}$ is given by
\begin{align}
\begin{aligned} \label{eq:free vpf}
Z_\text{hyper}(x,\tau,w) &= \frac{\left(\tau w x^\frac{3}{2};x^2\right)_\infty}{\left(\tau^{-1} w x^\frac{1}{2};x^2\right)_\infty} = \mathrm{PE}\left[\frac{\tau^{-1} w x^\frac{1}{2}-\tau w x^\frac{3}{2}}{1-x^2}\right] \\
&= Z_\text{vortex}^{1,1}(x,1,\tau,w).
\end{aligned}
\end{align}
The identity $Z^{1,1}_\text{vortex} = Z_\text{hyper}$ reflects the $\mathcal N = 4$ mirror symmetry, or equivalently the $\mathcal N = 4$ Seiberg-like duality, between the $U(1)$ theory with one flavor and the free twisted hypermultiplet theory. From \eqref{eq:large mass} and \eqref{eq:free vpf} we conclude that if we take the large mass flavor from $\{b_j\}$, the vortex partition function reduces to the product of the vortex partition function of the $U(N_c-1)$ theory with $N_f-1$ flavors and $Z_\text{hyper}$ of a free twisted hypermultiplet.

We have observed the two different limits of the vortex partition function. Using those results one can find a set of identities applicable to the $\mathcal N = 4$ Seiberg-like duality as shown in section \ref{sec:N=4}.

\acknowledgments

J. P. is supported in part by the National Research
Foundation of Korea Grants No. 2012R1A1A2009117, 2012R1A2A2A06046278, 2015R1A2A2A01007058.
J. P. also appreciates APCTP for its stimulating environment for research.
C. H. is grateful for discussions with Hee-Cheol Kim, Hyungchul Kim, Piljin Yi and Yutaka Yoshida.


\end{document}